\documentclass{aa}  
\usepackage{arydshln}
\usepackage{orcidlink}
\usepackage{hyperref}
\usepackage{booktabs}
\usepackage{multirow}
\usepackage{graphicx}
\usepackage{titlesec}
\usepackage[mathlines]{lineno} % mathlines adds numbering to equations too

\usepackage{txfonts}
\begin{document} 
%\linenumbers

\title{A predictive framework for realistic star–planet radio emission in compact systems}

   % \subtitle{I. Overviewing the $\kappa$-mechanism}

\author{J.~J.~Chebly\orcidlink{0000-0003-0695-6487}
          \inst{1}
          \and
          \
          C.~K.~Louis\orcidlink{0000-0002-9552-8822}\inst{2}
          \and 
          \ 
          A.~Strugarek\orcidlink{0000-0002-9630-6463}\inst{1}
          \and 
          \ 
          J.~D.~Alvarado~G\'omez\orcidlink{0000-0001-5052-3473}\inst{3}
          \
          \and 
          P.~Zarka\orcidlink{0000-0003-1672-9878}\inst{2} 
          }
          
   \institute{Universit\'e Paris Cit\'e, Universit\'e Paris-Saclay, CEA, CNRS, AIM, 91191 Gif-sur-Yvette, France\\
              \email{judy.chebly@cea.fr}
              \and
              LIRA, Observatoire de Paris, Universit\'e PSL, Sorbonne Universit\'e, Universit\'e Paris Cit\'e, CY Cergy Paris Universit\'e, CNRS, 92190 Meudon, France 
              \and
              Leibniz Institute for Astrophysics, An der Sternwarte 16, D-14482, Potsdam, Germany}
         
   \date{Received XX; accepted XX}

% \abstract{}{}{}{}{} 
% 5 {} token are mandatory
 
  \abstract{
  % Context heading (optional)
  % {} leave it empty if necessary  
  
  % Aims heading (mandatory)
  {Context: Radio emission from star–planet interactions~(SPI) beyond our solar system has yet to be firmly detected, primarily due to challenges such as weak signals, directional beaming effects, and low-frequency emissions that are blocked by Earth’s ionosphere. Addressing these obstacles calls for strategic target selection.}

  % Methods heading (mandatory)
  {Aim: This proof-of-concept study aims to improve SPI target prioritization by simulating SPI induced radio emission frequencies and estimating associated radio power to identify systems most likely to produce detectable signals.}

  % Results heading (mandatory)
  % {Method: We combine Zeeman–Doppler Imaging (ZDI) maps with 3D magnetohydrodynamic stellar wind simulations and use the ExPRES code to simulate SPI-driven radio emissions. This method is applied to systems like Tau~Boo, HD~179949, and HD~189733 to evaluate detectability with current and future radio telescopes.}
  {Method: We combine Zeeman--Doppler Imaging (ZDI) maps with 3D~magnetohydrodynamic (MHD) stellar wind simulations and use the ExPRES code to model SPI-driven radio emissions. We also estimate the intensity of these emissions using the Radio-Magnetic Scaling Law, based on the magnetic field and plasma density parameters from the 3D wind models. This approach is applied to systems such as Tau~Boo, HD~179949, and HD~189733 to assess their detectability with current and future radio telescopes.
    }

  % Conclusions heading (optional), leave it empty if necessary 
  {Conclusion: This framework, tested on benchmark systems, is applicable to any star–planet system with available ZDI maps and wind models. As magnetic field reconstructions and wind simulations improve, the method will become more robust. It provides a data-driven approach to prioritize targets and optimize telescope scheduling. This shall enable systematic exploration of magnetic SPI radio emissions across a wide range of exoplanetary systems.}
}

   \maketitle
%
%-------------------------------------------------------------------

\section{Introduction}

Magnetically driven radio emission in the solar system offers a compelling analogue for understanding star–planet interactions~(SPI) in exoplanetary systems. One of the most well-studied examples is the Jupiter–Io interaction, where Io's motion through Jupiter’s magnetosphere generates Alfv\'en waves that propagate along magnetic field lines, forming structures known as Alfv\'en wings~\citep{Bigg1964, Neubauer1980, Zarka1998, Paul2025}. These wings facilitate energy and momentum transfer between the moon and the plasma environment, accelerating electrons that emit intense, circularly polarized radio emission via the electron cyclotron maser instability~(ECMI). Similar emissions have also been observed from Jupiter’s interaction with Ganymede and Europa \citep{Louis2017, Zarka2018, Jacome2022}. This physical mechanism has been extended theoretically to close-in exoplanetary systems, where hot exoplanets orbit within a few stellar radii of their host stars~\citep{2001P&SS...49.1137Z, Zarka2007, Hess2011, Turnpenney2018}. 

\begin{figure}[hbt!]
    \centering
    \includegraphics[width=0.85\columnwidth]{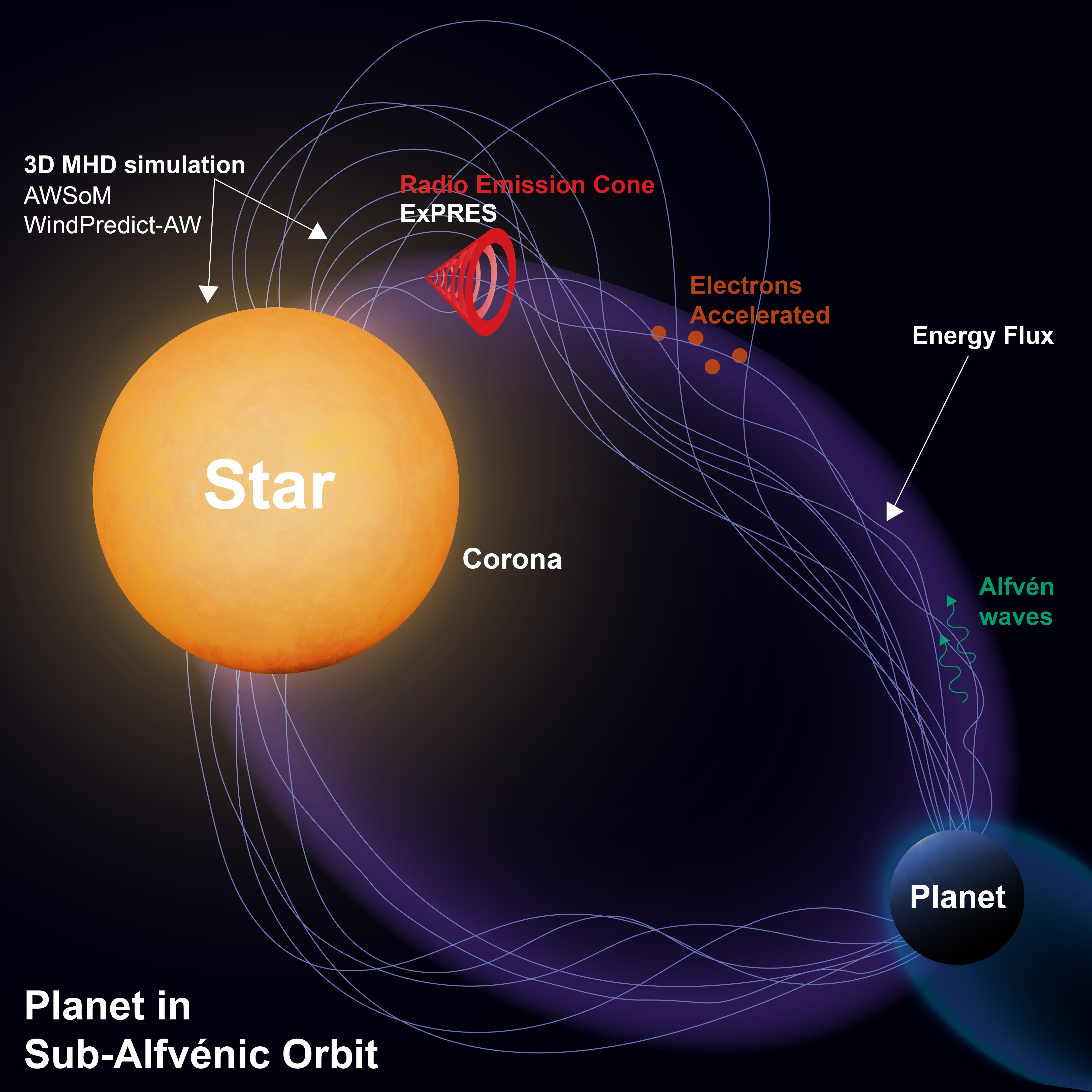}
     \caption{Artist's rendering illustrating star–planet interaction-induced radio emission. This illustration summarizes the physical processes driving radio emission, from the star’s activity to the direct energy flux generated by Alfv\'en~waves~(green color) electron acceleration at relativistic velocities~(red color circles), resulting in cyclotron radio emission~(red cone). The image also highlights the use of a 3D MHD stellar wind simulation, specifically the Alfv\'en Wave Solar Model (AWSoM) and WindPredict\_AW. These wind models are coupled with the ExPRES code to simulate and predict the SPI-induced radio emission.}
    \label{fig:SPI-radio emission}
\end{figure}

In this context, the structure of the stellar wind\footnote{Stellar winds are streams of charged particles continuously ejected from a star's outer atmosphere. They carry mass, momentum, and magnetic fields into space, shaping the space environment around stars.} plays a critical role. The wind sets the local plasma conditions (density, velocity, and magnetic field strength) which together determine the Alfv\'en speed and the location of the Alfv\'en surface~(AS)\footnote{This structure defines the boundary between the escaping stellar wind and the magnetically confined outflows, which do not contribute to the loss of angular momentum from the star.}, where the flow transitions from sub- to super-Alfv\'enic. Only within the sub-Alfv\'enic region, where the Alfv\'enic~Mach number\footnote{The Alfv\'enic Mach number $M_{\rm A} = v / v_{\rm A}$ compares the plasma flow speed $v$ to the Alfv\'en speed $v_{\rm A} = \dfrac{B}{\sqrt{\mu_0}\,\rho}$, where $\mu_0 = 4\pi \times 10^{-7}\,\mathrm{H\,m^{-1}}$ is the magnetic permeability of free space. Magnetic coupling is only possible in the sub-Alfv\'enic regime ($M_{\rm A} < 1$).}
~($M_\mathrm{A}) < 1$, can a planet maintain a magnetic connection to the star via Alfv\'en wings. These wings can channel energy back toward the stellar surface, potentially producing radio emission observable from Earth~(see Figure~\ref{fig:SPI-radio emission} and \citealt{Strugarek2015, Strugarek2016}). 

Despite strong theoretical support, detecting SPI-induced radio emission is a difficult objective that just starts to be met~\citep{Tasse2025}. 
The expected signals are inherently faint, often confined to low frequencies, and highly directional due to the beamed nature of ECMI emission. These factors, combined with absorption and scattering by the stellar wind and the interstellar medium, severely limit detectability~\citep{Dulk1985, Kavanagh2019, Weber2017b}. Moreover, there is Earth's ionospheric cut-off frequency~($<$10~MHz) requiring space-based instruments to probe this low-frequency regime. In addition, the intrinsic radio loudness of the host star presents a further obstacle. For example, in a continued radio-monitoring campaign of the G-dwarf HIP~67522, the star exhibited highly active radio behavior, stochastic variations, frequent bursts, and a high duty cycle of~69\%, all of which can readily mask any SPI-induced signals \citep{Ilin2025}.

Nonetheless, recent advancements in radio astronomy are revolutionising the study of SPI. 
Observatories such as the Low-Frequency Array~(LOFAR;~\citealt{vanHaarlem2013}), 
the Giant Metrewave Radio Telescope~(GMRT;~\citealt{Swarup1991, Gupta2017}), the Karl G.~Jansky Very Large Array~(JVLA;~\citealt{Perley2011}), 
the Five-hundred-meter Aperture Spherical radio Telescope~(FAST;~\citealt{Nan2006,Nan2011}),
the Australian Square Kilometre Array Pathfinder~(ASKAP;~\citealt{Johnston2008}), and such as NenuFAR~\citep{Zarka2020},
now provide unprecedented sensitivity and frequency coverage for stellar radio studies.

LOFAR, for example, revealed low-frequency~($\sim$150\,MHz) coherent emission from a quiescent M~dwarf~(GJ~1151), where 
the energy flux required to power this emission cannot be supplied by the star’s rotation-driven magnetic activity alone and may instead originate from a sub-Alfv\'enic interaction with a close-in planet~\citep{Vedantham2020}.

A LOFAR survey further uncovered 19 radio-emitting M~dwarfs, and for the most quiescent among them, 
emission characteristics suggest analogues of gas-giant magnetospheric processes, potential SPI signatures ~\citep{Callingham2021}. Beamformed LOFAR observations tentatively detected (3.2$\sigma$) bursts from the Tau~Boo system in the time-frequency plane~\citep{Turner2021}.
At higher frequencies, JVLA observations of YZ~Ceti detected coherent bursts aligned with the orbital phase of its innermost planet, consistent with a sub-Alfv\'enic SPI model, though stellar activity cannot yet be ruled out~\citep{Pineda2023}.  
Moreover, FAST observations of the flare-star AD~Leo revealed millisecond-scale fine structures in its radio bursts~(ranging from Jovian-like drifting stripes to blob-like solar analogues) that are consistent with electron cyclotron maser emission potentially driven by stellar flares or interactions with a planetary companion \citep{Zhang2023}. 
Recently, NenuFAR revealed a single, circularly polarized burst from the HD~189733 system exhibiting characteristics (polarisation, timing) consistent with cyclotron maser emission potentially induced by star–planet interaction. Though the authors caution that further observations are required to confirm a planetary origin~\citep{Zhang2025}.
Looking ahead, next-generation facilities like the Square Kilometre Array~(SKA), covering 50~MHz to 15~GHz will dramatically improve sensitivity and frequency coverage, enabling more detailed and comprehensive studies of these phenomena. 

Detecting SPI-induced radio emission offers a rare, indirect window into exoplanetary magnetic fields, key to understanding atmospheric protection and habitability~\citep{Yantis1977, Lazio2004, Zarka2007, Zarka2015}. These emissions also reveal properties of the stellar wind and magnetic interactions in star–planet systems~\citep{Lazio2024}. 

To guide and optimize observational searches for radio emission driven by SPI, we present a proof-of-concept study using a forward-modeling approach. While simulating SPI-induced radio emission is not new, existing tools such as MASER~\citep{Kavanagh2023} often rely on simplified assumptions and overlook complex magnetic topologies and plasma density variations. 
In this work, we couple the more sophisticated, electron-Cyclotron Maser Instability~(ECMI)-based, ExPRES code~(Exoplanetary and Planetary Radio Emission Simulator, \citealt{Louis2019a})
with state-of-the-art, data-driven 3D MHD stellar wind models and to the Radio-Magnetic Scaling Law (RMSL,~\citealt{Zarka2025HoE}) to produce realistic predictions of the amplitude and phase of radio emissions for specific star-planet systems. 

Using ExPRES, we simulate SPI-induced radio emissions in a way that remains physically consistent with the evolving star--planet geometry. As the planet orbits and the star rotates, changes in their relative positions and in the stellar wind conditions modify both the location of the radio source and the orientation of its emission beam. 

We note that the stellar wind models applied here are steady-state, capturing only spatial variability and not temporal. 
They cannot reproduce any variability on timescales shorter than the stellar rotation period (the reference timescale of the ZDI maps driving the models), nor on much longer timescales, where the static-field approximation of ZDI breaks down.
By accounting for the anisotropic nature of the emission, ExPRES predicts when and at what frequencies such signals would be detectable from Earth. 
In this study, the stellar magnetic field strength and topology are based on a single ZDI map~(snapshot), which allows to assess the morphology and strength of key temporal features of the emission for the epochs contemporary to these maps. This paves the way toward joint spectro-polarimetric and radio campaign leveraging our modelling framework for interpretation.

This paper is organized as follows: Section~\ref{sec:simulation setup} presents the simulation setup and input parameters, beginning with an overview of the stellar wind model, followed by a brief explanation of its coupling to ExPRES, and concluding with details of the radio emission modeling. Section~\ref{Prediction of radio emission} introduces the sample of stellar systems considered and presents the results of their simulated SPI-induced radio emissions. Section~\ref{sec:radio power} describes the estimation of radio power for each system and evaluates the detectability of the predicted signals by comparing them with the sensitivity limits of current and upcoming radio telescopes. Section~\ref{target-selecton} identifies the most promising detection targets among the systems studied, while Section~\ref{sec:caveats} discusses the key assumptions and limitations of our approach. Finally, Section~\ref{sec:conclusion} summarizes the main findings and outlines potential directions for future work.

\section{Radio emission prediction framework}
\label{sec:simulation setup}

\subsection{Stellar wind model}
\label{subsec:stellar wind model}

We have successfully coupled two 3D MHD stellar wind models with ExPRES: the Alfv\'en Wave Solar Model (AWSoM) and WindPredict-AW, each used independently.
AWSoM, part of the Space Weather Modeling Framework (SWMF; \citealt{Sokolov2013, vanderholst2014, Gombosi2018}), simulates the solar environment from the corona to the heliosphere using the BATSRUS MHD solver \citep{Powell1999}. It has been extensively applied in stellar wind and coronal mass ejection studies \citep[e.g.,][]{Alvarado2016,  Garraffo2017, Alvarado2018, Alvarado2019, Cohen2020, Kavanagh2021, Chebly2022, Chebly2023, Vidotto2023, Strickert2024}.
On the other hand, WindPredict-AW, built on the PLUTO MHD code \citep{Mignone2007}, has been employed to model both the solar wind~\citep{Varela2016,Reville2020, Parenti2022, Perri2024} and the wind of key exoplanet's host stars~\citep{Varela2018,  2022SpWea..2003164V, Strugarek2022, Reville2024, Pena2024, Colombo2024}. 

For this study, we focus on the AWSoM framework, as we are building upon the steady-state stellar wind simulations presented by \citet{Chebly2023}.
The magnetic inner boundary condition is set using surface magnetic field maps obtained from ZDI. These maps provide all the components of the surface magnetic vector field, but only the radial component is used in the simulations. This simplification is due to the model’s initial usage to model the solar corona, where the radial magnetic field dominates over the azimuthal and meridional components~\citep{Strugarek2022}.
The radial magnetic field from the ZDI map is analytically extrapolated using the potential field method\footnote{The potential field method assumes the coronal magnetic field is current-free ($\nabla \times \mathbf{B} = 0$), allowing it to be derived from a scalar potential. This yields a mathematically simple, force-free configuration, commonly used to approximate large-scale stellar magnetic fields.}, assuming the field becomes purely radial at a finite distance known as the source surface.

Moreover, all winds modeled in \citet{Chebly2023} share the same coronal base parameters, adopted from solar models: a base density of $n_0 = 2 \times 10^{11}~\mathrm{cm}^{-3}$, temperature $T_0 = 2 \times 10^6~\mathrm{K}$, Poynting flux per~unit magnetic~field~$S/B = 1.1 \times 10^6~\mathrm{J\,m^{-2}\,s^{-1}\,T^{-1}}$, and a transverse correlation length for Alfv\'en waves\footnote{The transverse correlation length controls the rate of turbulent dissipation that heats the corona and accelerates the stellar wind.} of $L_\perp \sqrt{B} = 1.5 \times 10^5~\mathrm{m}~T^{1/2}$. The stellar mass, radius, and rotation period are specified individually for each case. For a detailed description of the stellar wind model setup, see~\citet{Chebly2023}.

It should be noted that the stellar wind simulations themselves do not incorporate a planetary body. Instead, we approximate the planet’s position by selecting points along the orbit, from which the local magnetic field topology is extracted. 

We extract approximately 800 magnetic field lines intersecting the planetary orbit, 
with $\approx$400 with positive polarity and $\approx$400 with negative polarity. This procedure samples both parallel and anti-parallel directions to the local magnetic field without imposing any specific planetary magnetic configuration, thereby assuming that magnetic connectivity with the star can be established at the orbital location irrespective of the field polarity.

The plasma density is interpolated along each line. This data is then passed to the ExPRES code, allowing it to use plasma parameters extracted along selected magnetic field lines to generate predictions of radio emissions occurrence
~(see Appendix~\ref{app:coupling wind to ExPRES} for information on the input file format). Further details on the ExPRES modeling framework are provided in the following paragraph.

\subsection{Radio emission simulation: ExPRES}
\label{subsec:radio emission simulation}
% To predict radio emission from star–planet magnetic interactions, we use the ExPRES;~\citealt{Hess2011} simulation tool. 
The Exoplanetary and Planetary Radio Emissions Simulator (ExPRES) was originally developed to model Jupiter–Io decametric radio emissions generated by the Cyclotron Maser Instability (CMI,~\citealp{Hess2008a, Louis2017}). 
The code determines the visibility of radio sources for a given observer by self-consistently calculating the emission beaming geometry from the orientation and strength of the magnetic field, and the density and energy of the electrons.~It therefore requires 3D input data to properly account for these dependencies.

ExPRES, by itself, does not compute emission intensity because it does not model the CMI growth rate (but we provide amplitude estimates in this work in Section~\ref{sec:radio power}). Instead, it tracks the visibility of the emission cone relative to a given observer.
Later on ExPRES was extended to simulate CMI-driven radio emissions in a broader range of contexts,
including other solar system bodies such as auroral emissions on Saturn~\citep{Lamy2008a}, Saturn-Enceladus induced kilometric emissions~\citep{Lamy2013}, auroral emissions on Jupiter~\citep{Cecconi2012}, Europa- and Ganymede-Jupiter induced decametric emissions~\citep{Louis2017}, or to model radio emissions from exoplanetary systems involving SPI~\citep{Hess2011} and from stellar radio bursts~\citep{Zarka2025}.
In this study we will focus on SPI-induced radio emission.

ExPRES models radio emission as a hollow cone beam, with radio wave emitted along its edges. The properties of this beam depend on the local magnetic field strength and plasma density, as well as on the characteristics of the emitting electrons.
The maximum cyclotron emission frequency, $f_{\rm ce}^{\max}$, is set by the strongest radial magnetic field at the stellar surface, $B^{\max}$~(Eq.~\ref{eq:fce}). 
\begin{equation}
    f_{\rm ce}^{\max}~(\mathrm{MHz}) = 2.8 \times B^{\max}~(\mathrm{Gauss}).
    \label{eq:fce}
\end{equation}
At frequencies below this maximum, the shape of the emission cone varies. The opening angle $\theta$ depends on the electron velocity $v$ modified by the Lorentz factor $\Gamma^{-1} =\sqrt{1-\frac{v^2}{c^2}}$ (to take into account semi-relativistic effects), the speed of light $c$, the local emission frequency $f_{\rm ce}$~ and $f_{\rm ce}^{\max}$ and on $N$ the refractive index value~(Eq.~\ref{eq:beam_opening}). 
\begin{equation}
  \theta = \arccos \left( \frac{1}{N} \frac{\Gamma^{-1} v}{c}  \frac{1}{\sqrt{1-f_\mathrm{ce}/{f_\mathrm{ce}^\mathrm{max}}}}\right)
    \label{eq:beam_opening}
\end{equation}
Electron velocity distributions influence the cone width \citep[see, e.g.,][]{Hess2008a, Louis2019}: shell-like (horseshoe) distributions produce constant beams at $90^\circ$, while loss-cone distributions linked to Alfv\'enic acceleration generate narrower beams ($<90^\circ$ ).

The refractive index $N$ of Equation \ref{eq:beam_opening} deals with refraction effect inside the source: only if it tends to 1 the wave can propagate as a free wave. However, for certain characteristic frequencies of the plasma the refractive index may tend towards 0 (i.e., to a cutoﬀ frequency). It should be noted that when the $\arccos$ argument tends towards 1, the value of the beaming angle $\theta$ tends towards 0, in which case no emission is produced (the mechanism becoming theoretically ineffective below $\theta \sim$70$^\circ$, cf. \citealt{Pritchett1986}, where the value of $\theta$ drops very rapidly; see also Figure~4 in \citealt{Louis2019} for more details on the evolution of $\theta$ as a function of different parameters).

As we consider only pure circular polarization, ExPRES computes at each point of the user-defined radio source the wave frequency and the cutoﬀ frequency, and considers that emission is produced only if the wave frequency is above the cutoﬀ frequency \citep[see][for the exact computation of the refractive index $N$ inside the source]{Louis2019}.

\begin{figure}[hbt!] 
     \centering
      \includegraphics[trim={0cm 2cm 0cm 6cm},clip,width=0.85\columnwidth]{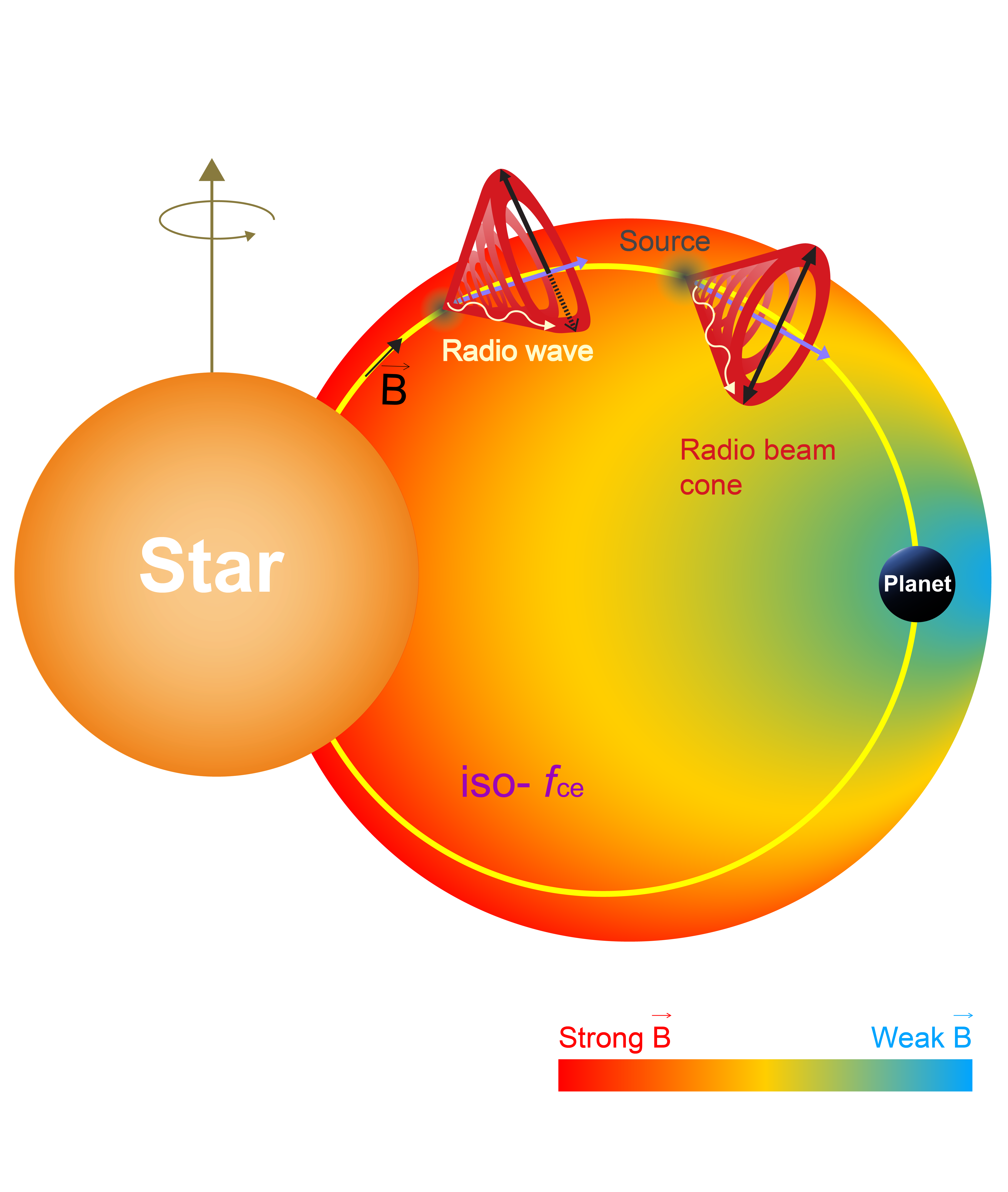}
      \caption{Conceptual illustration of radio wave propagation from the star toward the planet. The star is shown with a simplified dipolar magnetic field, and its rotation axis is indicated by the upward arrow. Two emission cones are depicted: the cone closer to the star lies in regions of stronger magnetic field (strong~$B$), while the outer cone is located in weaker fields (weak~$B$). The background color represents the magnetic field strength (orange–red: strong; green–blue: weak). As radio waves propagate into regions of stronger field (inner cone), the refractive index decreases, leading to reflection at the iso-$f_{\rm ce}$ surface and a flattening of the cone on that side. The illustration is not to scale.}
    \label{fig:iso-fce}
\end{figure}

Outside of the source, the refractive index differs from that inside the source as it is not symmetrical with respect to the magnetic field vector:~(i) either the wave is emitted towards a region where the norm of the magnetic field $B$ decreases, in which case it can continue its propagation without refraction effects~(Figure~\ref{fig:iso-fce}, cone closer to the planet);~(ii) or the wave is emitted towards a region where the norm of $B$ increases, reaching its cutoff frequency (i.e. $N \rightarrow 0$) at its iso-$f_\mathrm{ce}$, in which case it is reflected and the beaming angle $\theta$ decreases. This phenomenon happens close to the source, where the local cyclotron electron frequency $f_\mathrm{ce}$ is still close to that inside the source (and thus close to the wave frequency). 
In the code, the modification of the beaming angle is obtained from the Snell-Descartes law.  
This leads to flattening or narrowing of the emission cone on the side facing stronger fields, resulting in an asymmetric beam shape, as shown by the dashed black line in the cone near the star in Figure~\ref{fig:iso-fce}.

The simulations assume an optimistic scenario in which the star–planet interaction continuously generates radio emission along the connecting magnetic field lines, where local plasma and magnetic field conditions satisfy the resonance requirement~\citep{Louis2019}. We also define a fully specified star–planet–observer geometry, including the system’s inclination.
Maintaining temporal consistency is essential for planning observations and maximizing detection chances. However, this analysis is subject to the caveat that the underlying stellar magnetic field is assumed to be static, and thus long-term field evolution is not captured.

Furthermore, in the simulation we cover relatively long time intervals~(7~epochs/months) to explore the range of magnetic configurations that can arise for a given ZDI map. Clearly, running simulations over a seven-month timescale is not physically appropriate for our setup. However, this can be regarded as a gedanken experiment to explore how the signal may be affected by different ephemerides. In other words, extending the simulations over such a long duration effectively samples nearly all possible rotational-orbital configurations, thereby mapping out the range of pattern possibilities for a given ZDI~map and orbital distance.

For the electron energy distribution driving the emission, we adopt a loss-cone distribution with a characteristic energy of 20~keV. This choice is motivated by analogies with the Io--Jupiter system \citep{Piddington1977, Louis2020, Lamy2022, Louis2023, Mauduit2023}, and further supported by the analysis of CMI radio emissions from the AD~Leo system, which were also consistent with a 20~keV loss-cone population \citep{Zarka2025}. To assess the sensitivity of our results to electron energy, we additionally explore a wider range from 1~keV to 100~keV, encompassing plausible lower and upper limits (see Appendix~\ref{app:Entire Frequency-time diagram}). Moreover, we investigate the impact of the planet’s initial orbital phase on the predicted radio emission in Appendix~\ref{app:Ephemeris}.

\section{Prediction of radio emission occurrence for Tau~Boo, HD~179949 and HD~189733}
\label{Prediction of radio emission}

\subsection{Stellar sample}
\begin{figure}[hbt!]
    \centering
    \includegraphics[width=0.7\columnwidth]{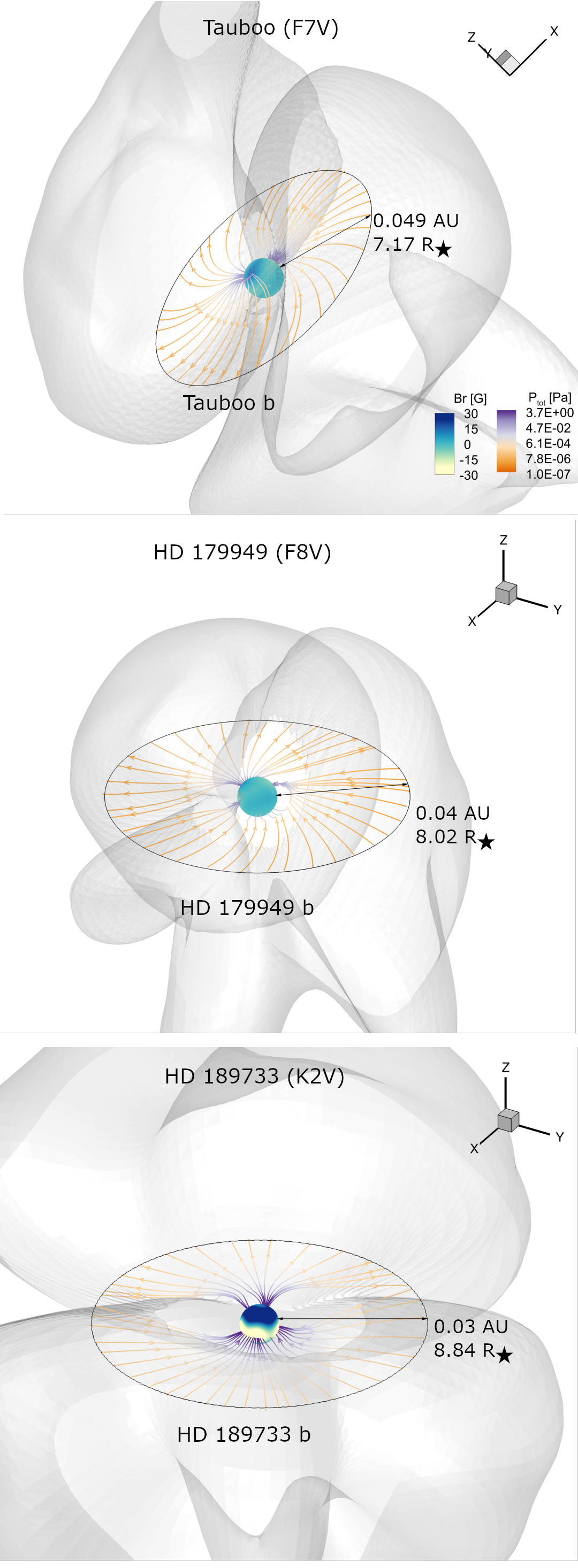}
     \caption{
    Simulated stellar wind environments for two F-type main-sequence stars (Tau~Boo and HD~179949) and one K-type star (HD~189733). Each panel shows the magnetic connectivity between the star and the planetary orbit: Tau~Boo (top), HD~179949 (middle), and HD~189733 (bottom). For Tau~Boo, the system is shown at its inclination relative to the observer, while for HD~179949 and HD~189733 a slightly tilted viewpoint is used to highlight the magnetic field topology. The stellar surface displays the radial magnetic field (in Gauss), reconstructed from ZDI maps, using a yellow-to-blue color scale. The translucent gray surface marks the Alfv\'en surface. For clarity, only a subset of magnetic field lines is shown, colored according to total wind pressure ($P_\mathrm{tot}$) from orange to purple, with darker tones indicating higher pressure. Both forward-connected (star to planet) and backward-connected (planet to star) field lines are included. The 3D axes (x, y, z) indicate the orientation of the stellar rotation axis, aligned with the z-axis.
}

    \label{fig:ZDI+AS+orbit}
\end{figure}

Our study is based on the sample of \cite{Chebly2023}, from which we select systems where the planet orbits within the predicted AS. These systems are Tau~Boo, HD~179949, and HD~189733. Figure~\ref{fig:ZDI+AS+orbit} shows the different stellar systems: Tau~Boo~(top), HD~179949~(middle), and HD~189733~(bottom). In each case, the stellar surface is color-coded by the radial magnetic field, and the AS is represented by the translucent gray surface. The planetary orbit (solid black line) lies inside the AS of the host star in all cases. The lines connecting the star to the planetary orbit represent magnetically selected field lines, color-coded by the total pressure (in pascals), with deeper purple indicating higher stellar wind pressure.

In the Tau~Boo system, the planets, companion star, and the primary star’s rotation are all tilted by roughly 45$^{\circ}$ relative to our line of sight, with components lying nearly in the same plane, indicating a well-aligned and stable configuration \citep{Justesen2019}.  
For HD~179949, the stellar inclination is estimated to be about $60^{\circ}$ with respect to our line of sight~(given $v \sin i$ of 7\,km/s, \citealt{Valenti2005}), while the planet’s orbital inclination is $67.7 \pm 4.3^{\circ}$~\citep{Brogi2014}.

The HD~189733 system is observed nearly edge-on, with a planetary orbital inclination $i \simeq 85.7^\circ$ \citep{Stassun2017}, allowing the planet to transit the star. Rossiter-McLaughlin measurements for HD 189733~b yield a sky-projected spin-orbit angle of $\lambda = -1.4 \pm 1.1^{\circ}$, indicating close alignment between the stellar rotation axis and planetary orbit, with both observed nearly equator-on \citep{Winn2006}.

Stellar and planetary parameters for these systems are summarized in Table~\ref{tab:star_properties}, with stellar properties listed in the upper section. The ZDI maps correspond to the following epochs: Tau~Boo (2010; \citealt{Fares2013}), HD~179949 (2009; \citealt{Fares2012}), and HD~189733 (2006; \citealt{Fares2010}).

\begin{table}[h]
\centering
\caption{Stellar and planetary parameters for Tau~Boo, HD~179949, and HD~189733 systems.}
\label{tab:star_properties}
\resizebox{\columnwidth}{!}{%
\begin{tabular}{p{5cm}ccc}
\hline\hline
\multicolumn{4}{c}{\textbf{Stellar parameters}} \\
\hline
Parameter & Tau~Boo & HD~179949 & HD~189733 \\
\hline
Spectral type & F7V & F8V & K2V\\[3pt]
Rotation period (days) & 3 & 7.6 & 12.5 \\[3pt]
Mass ($M_\odot$) & 1.34 & 1.21 & 0.82 \\[3pt]
Radius ($R_\odot$) & 1.46& 1.19 & 0.76 \\[3pt]
Distance (pc) & 15.6 & 27.5 & 19.8 \\
\hline
\multicolumn{4}{c}{\textbf{Planetary parameters}} \\
\hline
Mass ($M_J$) & 5.95 & 0.916 & 1.13 \\[3pt]
Radius ($R_J$) & 1.06& 1.2 & 1.1 \\[3pt]
Orbital period (days) & 3.3 & 3.09 & 2.2 \\[3pt]
Semi-major axis (AU) & 0.048& 0.04$^{\rm +0.00045}_{\rm -0.00046}$& 0.03153$\pm$0.00011\\[3pt]
Eccentricity & 0.0114$\pm 0.006$ & $0.016^{\rm+0.017}_{-0.011}$ & $0.027^{+0.021}_{-0.018}$ \\

\hline
\end{tabular}%
}
\tablefoot{
The upper part, ``Stellar Parameters'', lists host star properties, including rotation period, mass, radius, and distance taken from \cite{Chebly2023} and references therein. The lower part, ``Exoplanet Parameters'', provides planetary properties: radii and masses are from \citet{Rosenthal2021} for Tau~Boo~b, \citet{Brogi2012} for HD~179949~b, and \citet{Kokori2023} for HD~189733~b. Orbital periods are assumed equal to planetary rotation periods, based on the assumption that close-in planets, such as hot Jupiters, are tidally locked. Eccentricities are taken from \citet{Stassun2017,Rosenthal2021}, and semi-major axes from \citet{Rosenthal2021} for Tau~Boo~b and HD~179949~b while for HD~189733~b is taken from \citet{Kokori2023}.}
\end{table}

\titlespacing*{\subsection}{0pt}{6pt}{1pt} % 
\subsection{Time-frequency diagram}
In this section, we display and analyse the frequency–time pattern and its evolution across different simulated months for each star system. Although the radio emission due to SPI was simulated over seven months, we present only three representative epochs that best illustrate the frequency variability. The results of the full set of simulations for different electron energies are provided in Appendix~\ref{app:Entire Frequency-time diagram}.

\titlespacing*{\subsection}{0pt}{6pt}{1pt} % 
\subsection*{Tau Boo}
\label{subsubsec:tauboo}

\begin{figure}[hbt!] 
     \centering
      \includegraphics[width=0.85\columnwidth]{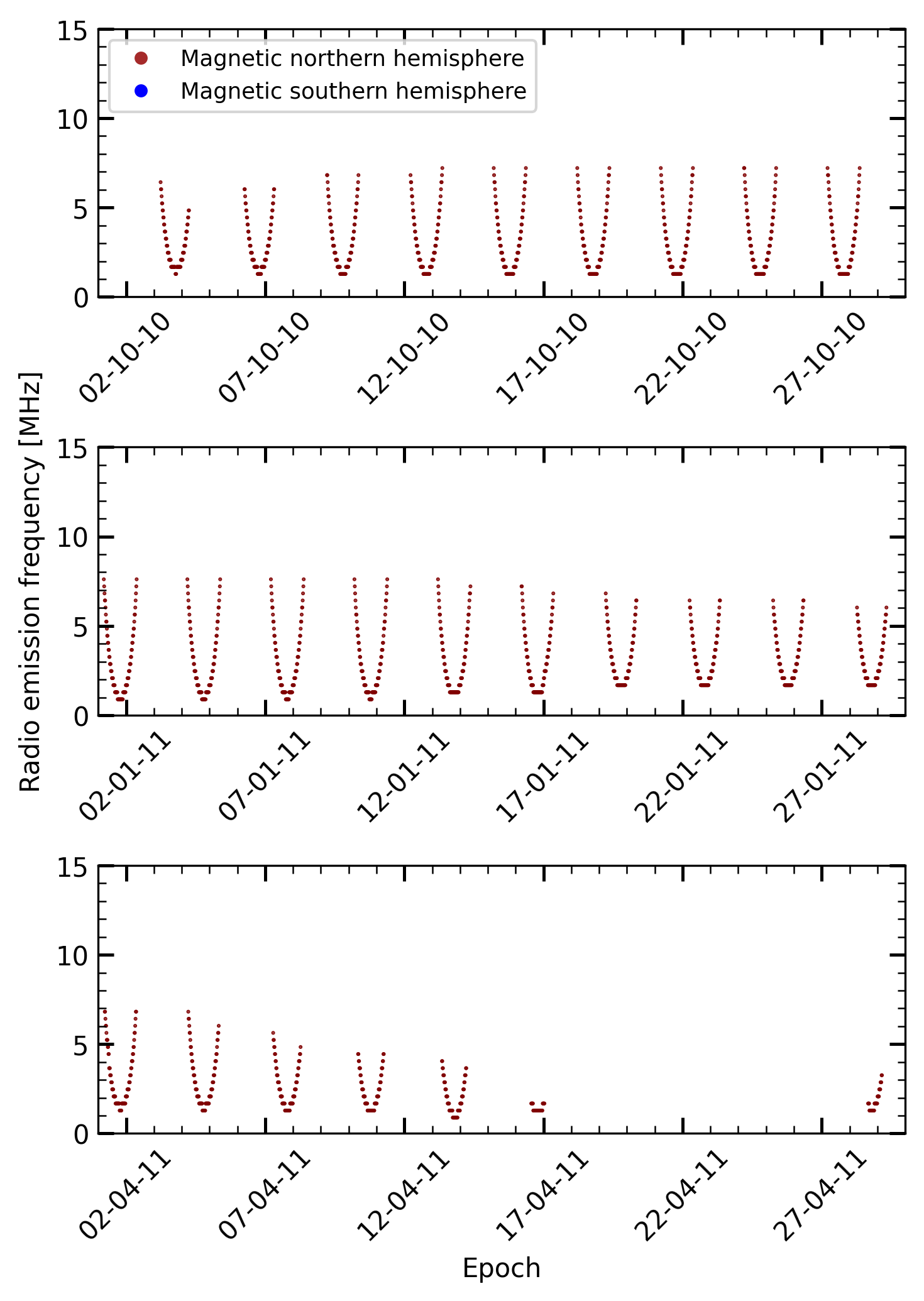}
      \caption{Time–frequency spectrograms showing the simulated radio emission from star–planet magnetic interactions in the Tau~Boo system for selected months; October~2010 (top), January~2011 (middle), and April~2011 (bottom). The y-axis denotes the radio frequency, while the x-axis represents time, with each tick corresponding to a single day within a given month (formatted as ``day:month:year''). Frequencies are color–coded by the degree of circular polarization: blue indicates emission from the magnetic southern hemisphere, 
    corresponding to field lines directed from the star to the planet, while red indicates emission from the magnetic northern hemisphere. In both cases, the color refers to the polarity of the field at the stellar surface (or carried by the wind).}     
    \label{fig:tauboo_frequency}
\end{figure}

Figure~\ref{fig:tauboo_frequency} shows the simulated SPI-induced radio emission of Tau~Boo for the epochs of October~2010~(top), January~2011~(middle), and April~2011~(bottom). 
Tau~Boo’s simulated radio emission spans 0.1--8~MHz across the three epochs, falling below Earth’s ionospheric cutoff ($\sim$10~MHz) and thus remaining undetectable from the ground for the ZDI map considered in this study. 

An increase in the stellar magnetic field strength raises the cyclotron frequency linearly~($f_{\rm ce} \propto B$). Stronger magnetic fields shift the emission to higher frequencies. 
At the same time, the coronal and wind densities increase, which raises the plasma frequency more slowly, $f_{\rm pe} \propto \sqrt{n_e}$. Because of this difference in scaling, the ratio $f_{\rm pe}/f_{\rm ce}$ generally decreases as $B$ increases, unless the density rises very steeply (faster than $B^2$). The plasma cutoff frequency is primarily governed by the local electron density rather than directly by the magnetic field strength. 
In stellar atmospheres, regions of strong magnetic field generally coincide with higher densities below the transition region, where the two are correlated. 
Above the transition region, where the density decreases more gradually, stronger magnetic fields can instead facilitate more favorable conditions for both the generation and escape of CMI emission.

The maximum emission frequency gradually decreases from October~2010 to April~2010.
Moreover, a consistent feature across all epochs is the absence of detectable emission from the magnetic south~(i.e. field lines pointing from star to planet), due to the system's inclination ($\sim44.5^\circ$), which favors detectable emission from the magnetic northern hemisphere. 
This assumes that the visible pole maintains its polarity 
(although in systems such as Tau Boo this polarity can reverse, potentially on timescales shorter than those simulated here).
Since the emission is beamed in a specific direction that is determined by the local frequency and resonant electron velocity~(see Eq.~\ref{eq:beam_opening}), the observer's viewing angle is crucial. Thus, non-detections, such as during April~18–27, 2011~(Figure~\ref{fig:tauboo_frequency}, bottom panel), are likely due to the emission beam pointing away from Earth rather than the absence of emission.

The radio emission cone, anchored to the stellar magnetic field, follows the footpoint path, which may or may not co-rotate with the planet depending on the magnetic topology, producing a periodic modulation
(either at the exoplanet period, or at the synodic period between the exoplanet and its host star~\citealt{Louis2025}) of the signal as seen by an Earth-based observer\footnote{For a dipolar field of arbitrary inclination, this modulation occurs either at the exoplanet's orbital period if the dipole is axisymmetric (Saturn-like case), or at the synodic period between the exoplanet and its host star) if the dipole is tilted, offset, or the magnetic field is more complex (Jupiter-like cases).}.
For Tau~Boo, this leads to a recurrence of the emission approximately every 3.5~days, in line with the planet’s orbital period. The resulting emission exhibits a characteristic ``U''-shaped frequency–time pattern, which is strongly influenced by the observer’s magnetic latitude.
In the case of Tau~Boo, the observer is positioned at a fixed magnetic latitude of approximately $45^\circ$, corresponding to a mid/high-latitude viewing geometry. This configuration favors a “O”-shaped pattern in the time-frequency diagram. 

\begin{figure}[hbt!] 
     \centering
      \includegraphics[width=\columnwidth]{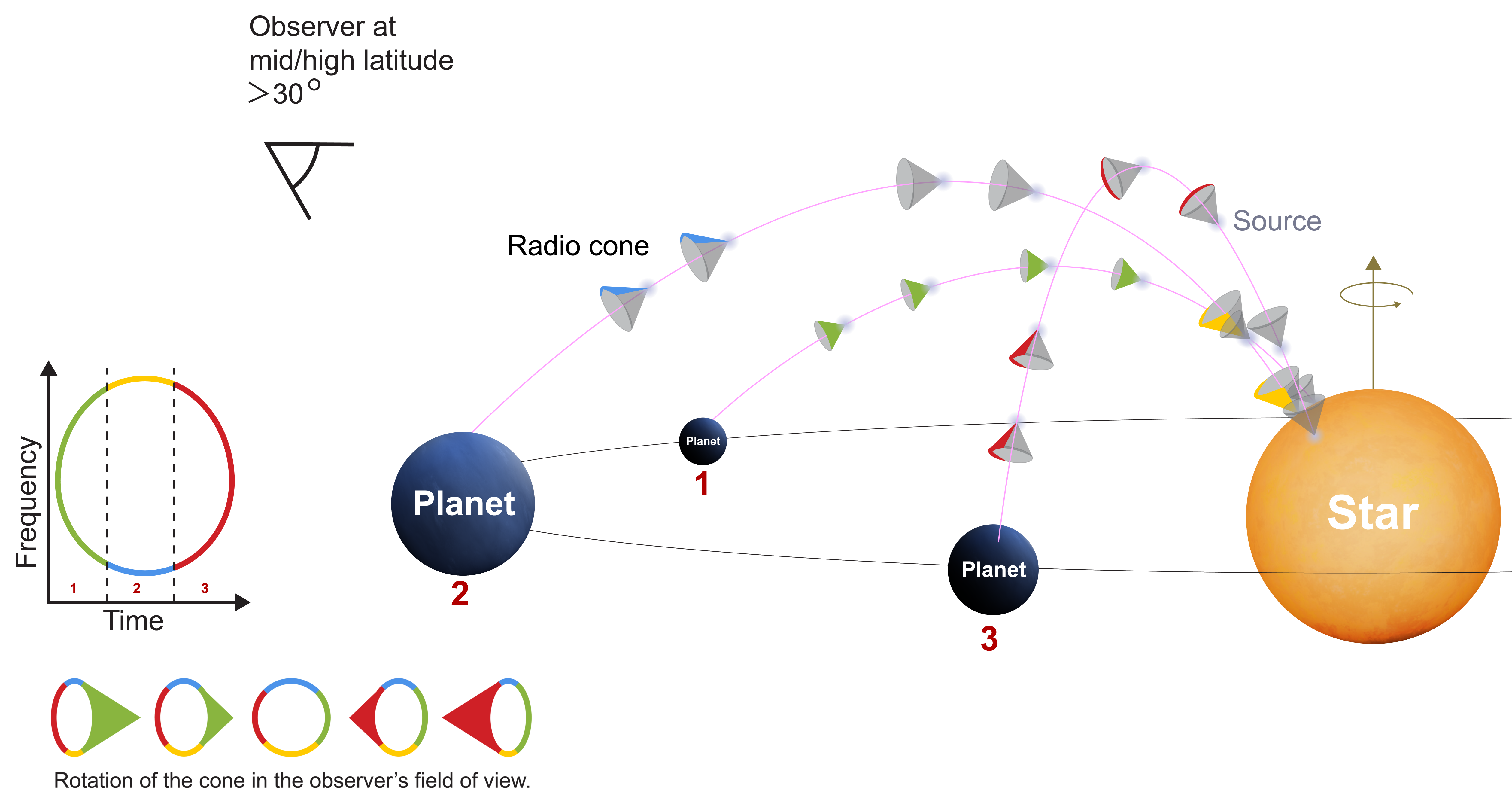}
      \caption{A simple illustration of how the observed radio frequency pattern changes as the star–planet system rotates, assuming a dipolar, axisymmetric magnetic field. We show the frequency pattern as seen from an observer located at mid/high magnetic latitudes from the star system. 
      The planet is shown moving from positions 1 to~3. It appears larger when it is closer to the observer (position~2), and the same perspective effect applies to the radio beam cones. Colored cones indicate the radio emission directed toward the observer: blue/yellow (top/bottom), green/red (right/left), and gray indicates directions with no emission toward the observer (represented by the eye symbol). As the planet orbits, different portions of the emission cone become visible to the observer, leading to changes in the observed frequency pattern. The magnetic field line is shown in magenta color. The illustration is not to scale.}
      \label{fig:mid/high-latitude_pattern}
\end{figure}

Figure~\ref{fig:mid/high-latitude_pattern} illustrates the expected frequency–time pattern of an observer at high/mid latitude using an idealized axisymmetric dipole model, neglecting refraction effects. For simplicity and clarity, we illustrate only two representative radio cones at three specific locations along a magnetic field line, although in reality, ECMI-driven radio emission sources are distributed along the field line to regions where cyclotron frequency~($f_{\rm ce}$) $>>$ plasma frequency~($f_{\rm pe}$), and an unstable electron distribution exist. 

We remind the reader that the emission frequency follows the local cyclotron frequency, $f_{\mathrm{ce}}~[\mathrm{MHz}] \simeq 2.8\,B~[\mathrm{G}]$. Along a given star–planet flux tube in the SPI scenario, $|B|$ increases toward the stellar footpoint; consequently, emission mapped closer to the star occurs at higher frequencies, 
whereas emission nearer the planet is at lower frequencies (along the same line). This ordering is not altered by atmospheric escape: enhanced density primarily affects the CMI growth and escape via the condition $f_{\mathrm{pe}} \lesssim f_{\mathrm{ce}}$ and may shift or suppress the source, but it does not affect the relation $f_{\mathrm{ce}} \propto B$. By contrast, if we consider radio emissions originating from the exoplanet itself, then the emission will be linked to the exoplanetary magnetic field, such that regions closer to the exoplanet correspond to higher frequencies, while emission farther from the planet (i.e., slightly closer to the star but still within the exoplanet magnetosphere) will occur at lower frequencies.

\begin{figure}[hbt!] 
    \centering
    \includegraphics[width=0.8\columnwidth]{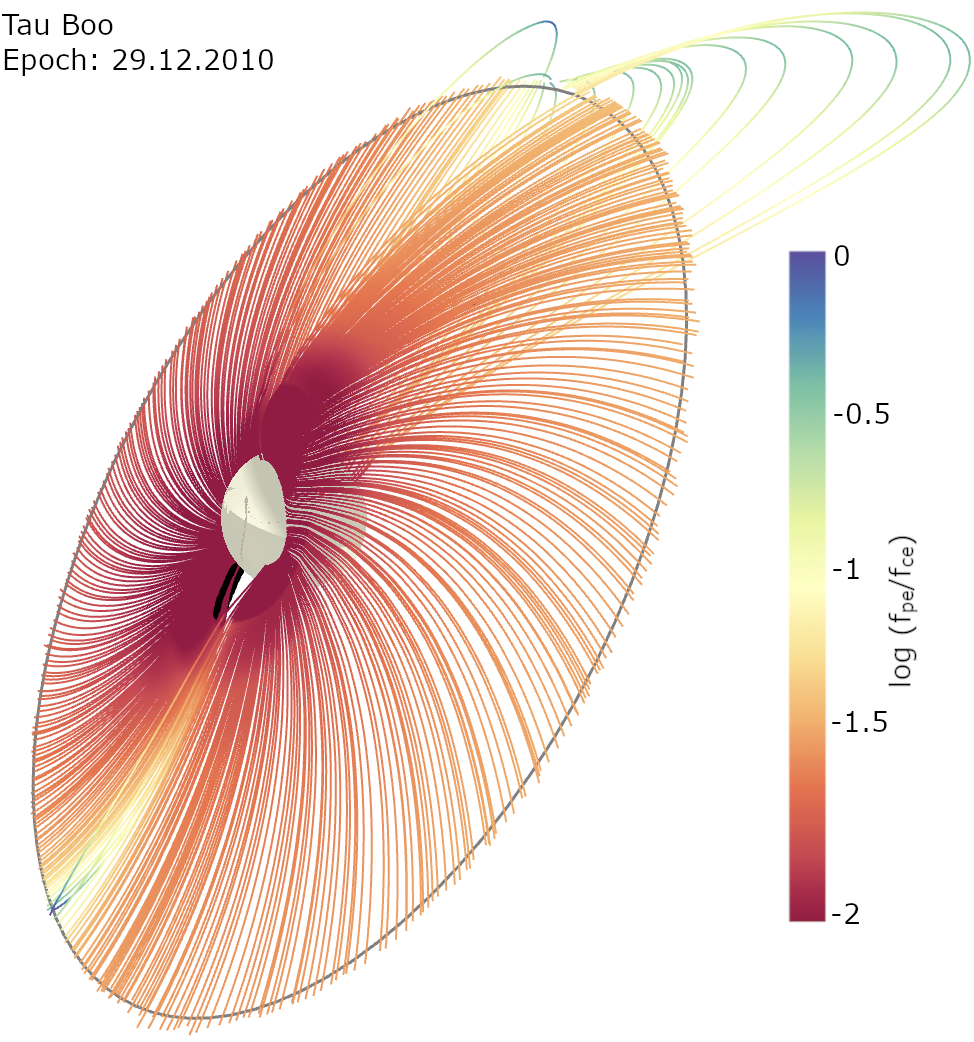}
   \caption{Stellar wind simulation of the Tau~Boo system, with the star shown in gray and the planet’s orbit as a solid black line. Magnetic field lines connecting the star to the orbit are color-coded by the logarithmic ratio of plasma frequency ($f_{\rm pe}$) to cyclotron frequency ($f_{\rm ce }$); $\log (\frac{f_{\rm pe}}{f_{\rm ce}}) < - 1$ indicates that radio waves can escape, while $\log (\frac{f_{\rm pe}}{f_{\rm ce}}) > - 1$ indicates reflection or absorption near the star. The system is displayed at an inclination of $44.5^{\circ}$ relative to Earth, and black dots mark radio source positions at 29 December 2010. An interactive version showing evolving source positions is available at \href{https://judy94.github.io/Star-system-FP-FC-visualization/Tauboo_fpfce_source_corrected.html}{``Tau~Boo system source position animation''}.In the latter, you can rotate and zoom to examine the source position in more detail. Note that the animation must be paused before rotating the view.}
    \label{fig:Tauboo-fpfc}
\end{figure}

In Figure~\ref{fig:mid/high-latitude_pattern}, at position~(1), the observer detects low- to mid-frequency emission (green arc) originating from regions near the planet and near the apex of the magnetic field, where the magnetic field strength is relatively weak. At position~(2), the cone is oriented such that both high-frequency emission from regions closer to the star~(yellow arc), where the magnetic field is stronger, and low-frequency emission from near the planet~(blue arc)~could be visible. At position~(3), as the emission cone begins to move out of the observer’s line of sight, the observer primarily detects low to mid-frequency emission (red arc) from near the planet and near the apex of the magnetic field, symmetric to (1).

Although a mid- to high-latitude viewing angle would favor a symmetric ``O''-shaped emission pattern, the Tau~Boo simulation shows only the lower portion. The missing upper part (yellow arc) results from the realistic, non-dipolar stellar magnetic field and the associated plasma density, which modify the refractive index in Equation~\ref{eq:beam_opening}. High-density regions prevent some emissions from escaping, and refraction along the iso-$f_{\rm ce}$ surfaces further affects propagation. Since $f_{\rm pe}$ increases faster than $f_{\rm ce}$ toward the star, higher-frequency emissions are more strongly influenced by these effects.  

Figure~\ref{fig:Tauboo-fpfc} illustrates this behavior, showing the logarithmic ratio of $f_{\rm pe}$ to $f_{\rm ce}$ along magnetic field lines and the radio source positions at epoch 29.12.2010 (black circles, confined to a single hemisphere). ECMI becomes inefficient when $f_{\rm pe}/f_{\rm ce} \gtrsim 0.1$ ($\log f_{\rm pe}/f_{\rm ce} \gtrsim -1$; \citealt{Hilgers1992,Zarka2001}). The colormap spans $\log f_{\rm pe}/f_{\rm ce}$ from $-2$ to $0$, with orange-red regions favorable for ECMI and yellow-blue regions indicating suppression or emission trapping. Tau~Boo thus exhibits a combination of both regimes, explaining why only part of the idealized emission pattern is visible.
 
\subsection*{HD 179949}
\label{subsubsec:HD179949}
\begin{figure}[hbt!] 
     \centering
      \includegraphics[width=0.9\columnwidth]{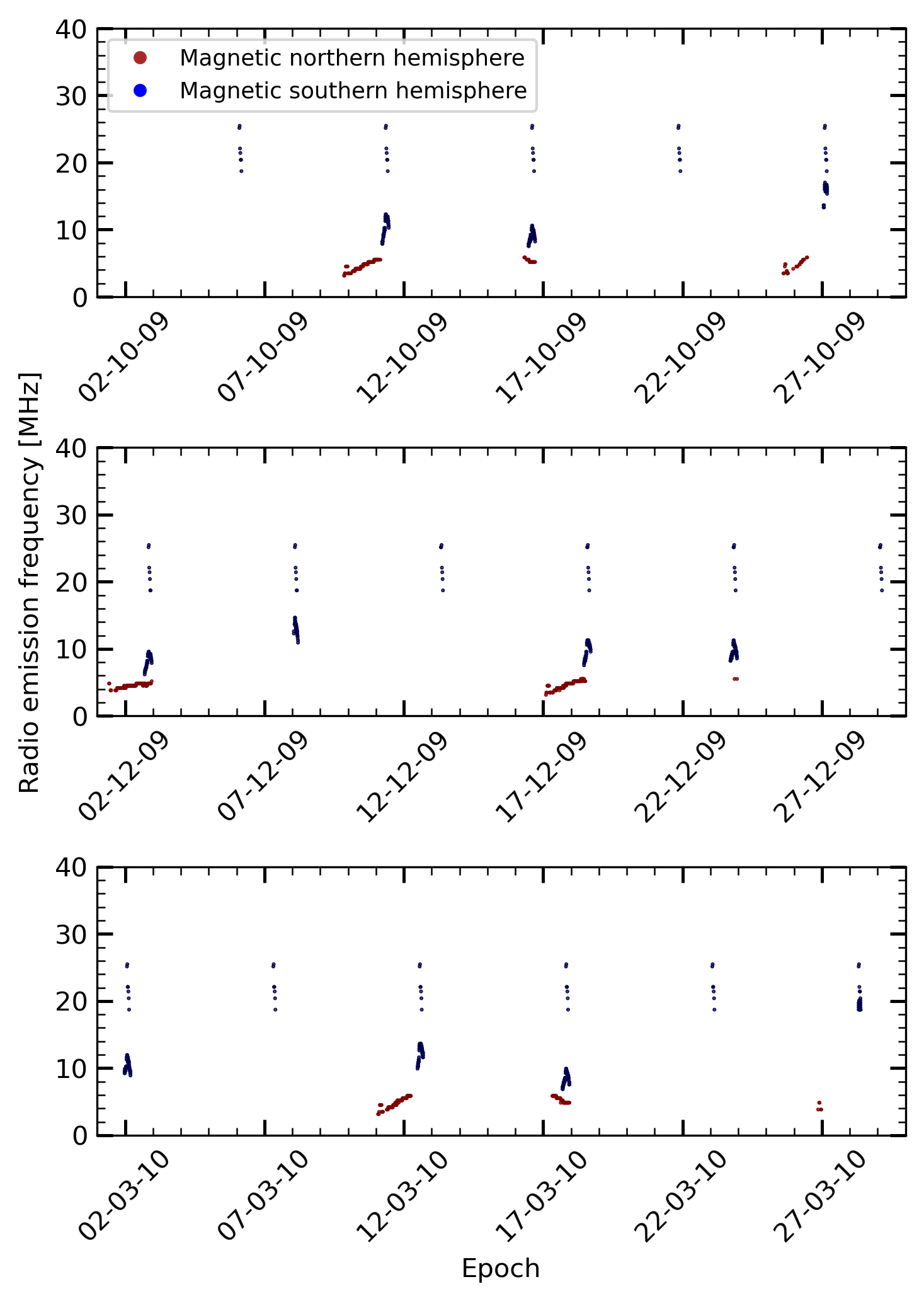}
      \caption{Same caption as Figure~\ref{fig:tauboo_frequency}, but for HD~179949. We show three simulated epochs: October~2010~(top), January~2011~(middle), and April~2011~(bottom).}
      
    \label{fig:HD179949_frequency}
\end{figure}

Figure~\ref{fig:HD179949_frequency} shows the simulated SPI-induced radio emission from HD~179949 for the epochs of October 2009~(top), December~2009~(middle), and March~2011~(bottom). The simulations cover frequencies ranging from 0.1 to approximately 30~MHz and span the period from 2009 to 2010, including the observed epoch in October 2009. 

Unlike Tau~Boo, HD~179949 exhibits radio emission originating from both magnetic hemispheres (north and south), featuring prominent high-frequency components near 30~MHz as well as low-frequency emission below 8~MHz.
Emissions from the magnetic southern hemisphere~(in blue) reach higher frequencies than those from the magnetic northern hemisphere~(in red); the former reach almost 30~MHz, but rarely go below 15~MHz, while the latter do not exceed 8~MHz. 
% This indicates a higher magnetic field strength in the southern hemisphere.
%at those frequencies. 
The radio emission varies on a timescale of approximately 3-4~days, which is comparable to the exoplanet orbital period of 3.09~days. Overall, the highest cyclotron frequency observed remains below 30~MHz, which lies above Earth's ionospheric cutoff frequency, making it potentially detectable by ground-based instruments.

Moreover, radio emissions from the magnetic south are typically sporadic and appear as vertical-line-like features in the frequency–time plots (see blue-colored signals in Figure~\ref{fig:HD179949_frequency}). In contrast, the magnetic north produces more elongated patterns along the time axis, which correspond to the low-frequency blue arc visible in Figure~\ref{fig:mid/high-latitude_pattern}.

\begin{figure}[hbt!] 
    \centering
    \includegraphics[width=0.9\columnwidth]{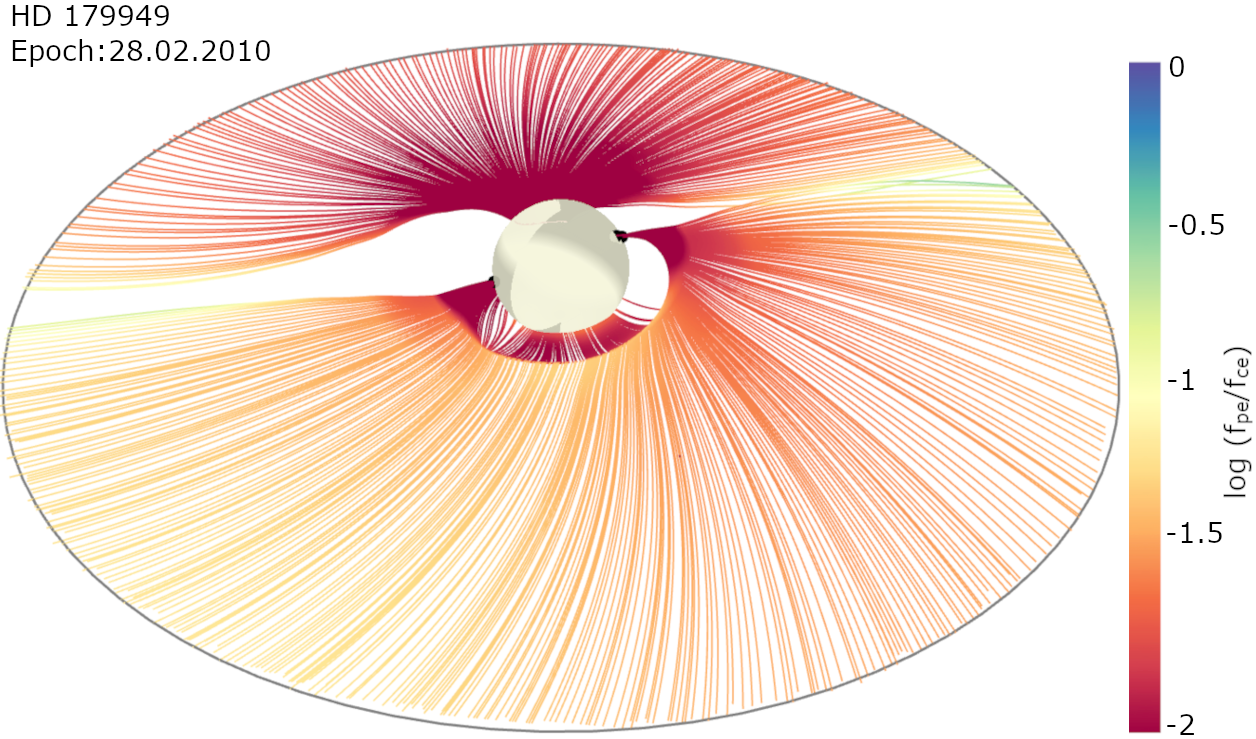}
    \caption{Same caption as Figure~\ref{fig:Tauboo-fpfc}, but for the case of HD~179949 at the epoch of 28 February 2010. The reader is invited to view the interactive version at \href{https://judy94.github.io/Star-system-FP-FC-visualization/HD179949_fpfce_source.html}{``HD~179949 system epoch animation''}.}
    \label{fig:HD179949-fpfc}
\end{figure}

Propagation effects, governed by local plasma conditions, further shape the observed emission. In the simulation, the refractive index (Equation~\ref{eq:beam_opening}) depends on the local plasma density and magnetic field~(Figure~\ref{fig:HD179949-fpfc}). Near the star, where $f_{\rm ce} \lesssim f_{\rm pe}$ (Section~\ref{subsec:radio emission simulation}), radio waves cannot always escape as free waves. Emission directed toward regions of increasing magnetic field can also be reflected along iso-$f_{\rm ce}$ surfaces, so parts of the emission cone remain unobservable, as indicated by the uncolored regions in Figure~\ref{fig:mid/high-latitude_pattern}. These propagation effects explain why only portions of the expected O-shaped emission pattern are visible in the magnetic north.
Similarly, the vertical structure of magnetic northern and southern emissions (blue in Figure~\ref{fig:HD179949_frequency}) shows that neither higher- nor lower-frequency emissions reach the observer.  
% Both the variations in this angle and the planet's orbital period can cause arcs to appear more stretched or distorted.

\subsection*{HD 189733}
\label{subsubsecc:HD189733}

\begin{figure}[hbt!] 
     \centering
      \includegraphics[width=0.85\columnwidth]{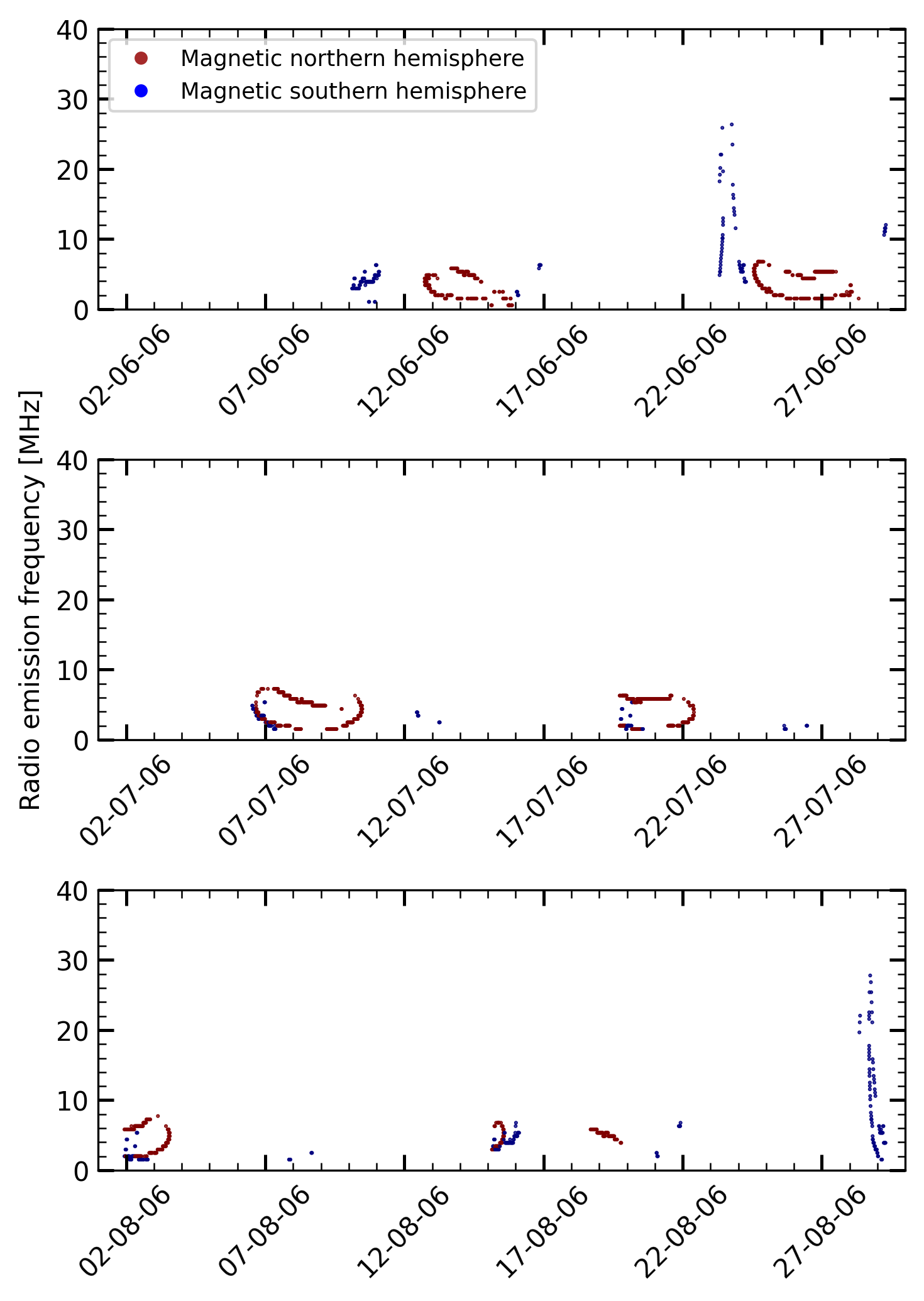}
      \caption{Same caption as Figure~\ref{fig:tauboo_frequency}, but for HD~189733. We show three simulated epochs: June, July, and August 2006 from top to bottom.}
    \label{fig:HD189733_frequency}
\end{figure}

In Figure~\ref{fig:HD189733_frequency}, we present the simulated radio emission frequencies for HD~189733 during the months of June, July, and August 2006 (from top to bottom). Each epoch reveals distinct features in the radio emission pattern. The computed frequencies span the range 0.1--30~MHz, exceeding the 10~MHz ionospheric cutoff of Earth's atmosphere, and are therefore potentially detectable from the ground-based radio telescopes.

The radio emission from HD~189733 occurs approximately every 10 days, compared to every $\sim 3-4$ days in HD~179949. This reflects the differences between the stellar rotation and planetary orbital periods. In HD~189733, the star's rotation period is nearly 5 to 6 times as long as the planet's orbital period. This leads to longer intervals between successive interaction phases and hence more widely spaced radio emission.

The time–frequency diagram for HD~189733 shows contributions from both the magnetic northern and southern hemispheres, similar to HD~179949. However, in this system, the magnetic north generates pronounced arc-shaped patterns that evolve over a highly compressed timescale. The appearance of these arcs is strongly influenced by the observer’s latitude.
% \textcolor{purple}{In this system, however, the planet orbits 5 to 6 times slower than the star, which stretches the features and makes them appear more elongated along the time axis.}

\begin{figure}[hbt!] 
     \centering  \includegraphics[width=\columnwidth]{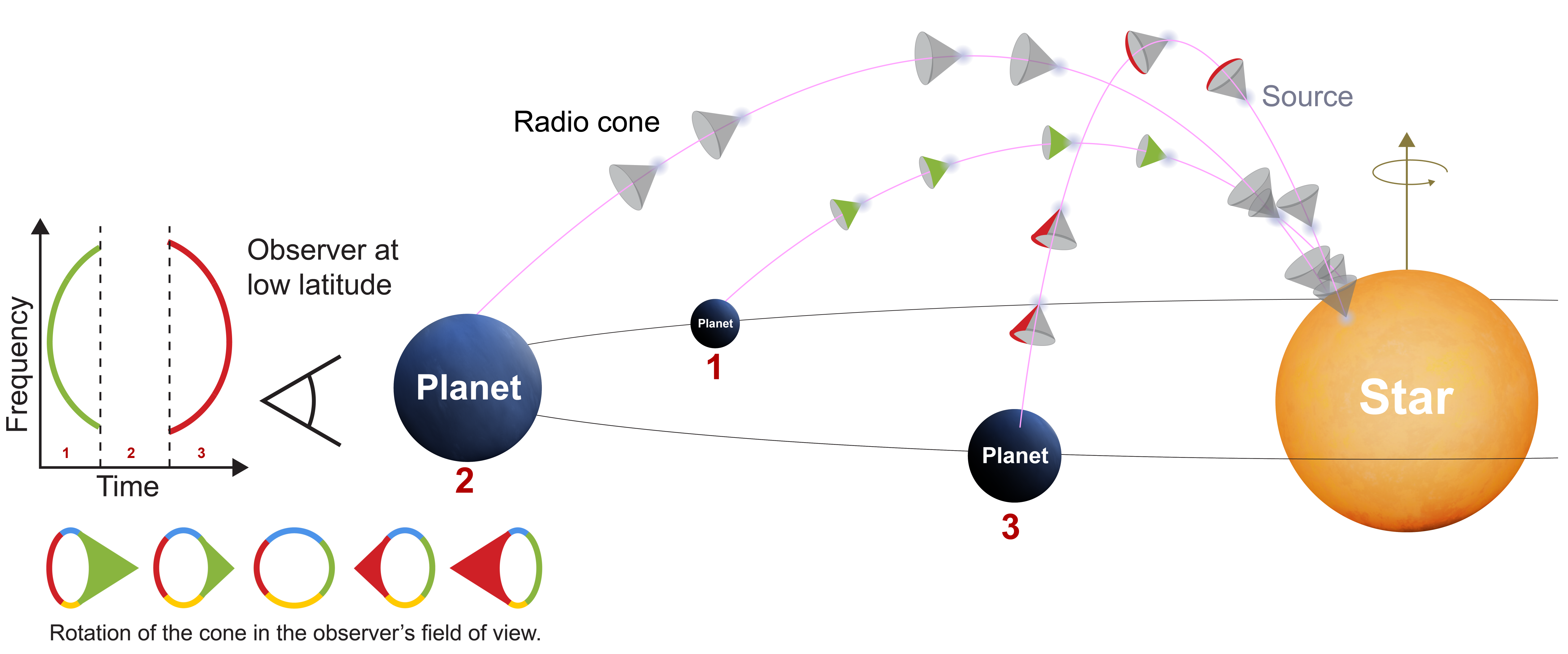}
      \caption{Same caption as Figure~\ref{fig:mid/high-latitude_pattern} but for the case of a low magnetic latitude observer.}
      \label{fig:low_latitude_pattern}
\end{figure}

For a low-latitude observer (Figure~\ref{fig:low_latitude_pattern}), the emission along an active magnetic field line can be separated into two phases during its rotation: (1) a low- to mid-frequency signal (green arc) as the emission cone first crosses west of the meridian, and (2) a second low- to mid-frequency signal (red arc) as it crosses east of the meridian.
In HD~189733, the observed radio emission directly reflects these two cases, forming bracket-like structures resembling an open parenthesis~(C-shape) and closed parentheses~(inverted C-shape). This occurs because the stellar rotation period is roughly 5-6 times longer than the planet’s orbital period. As the planet orbits the star, it moves from one magnetic field line to another, producing apparent shifts in the location of the associated interaction regions. As a result, the emission does not appear as continuous arcs or ``O'' shape, but as separated features in time. Case 1 corresponds to the C-shape (opening parenthesis of the inverted U), while case 2 corresponds to the inverted C-shape (closing parenthesis), appearing about half a rotation later. 

\begin{figure}[hbt!] 
    \centering
    \includegraphics[width=0.9\columnwidth]{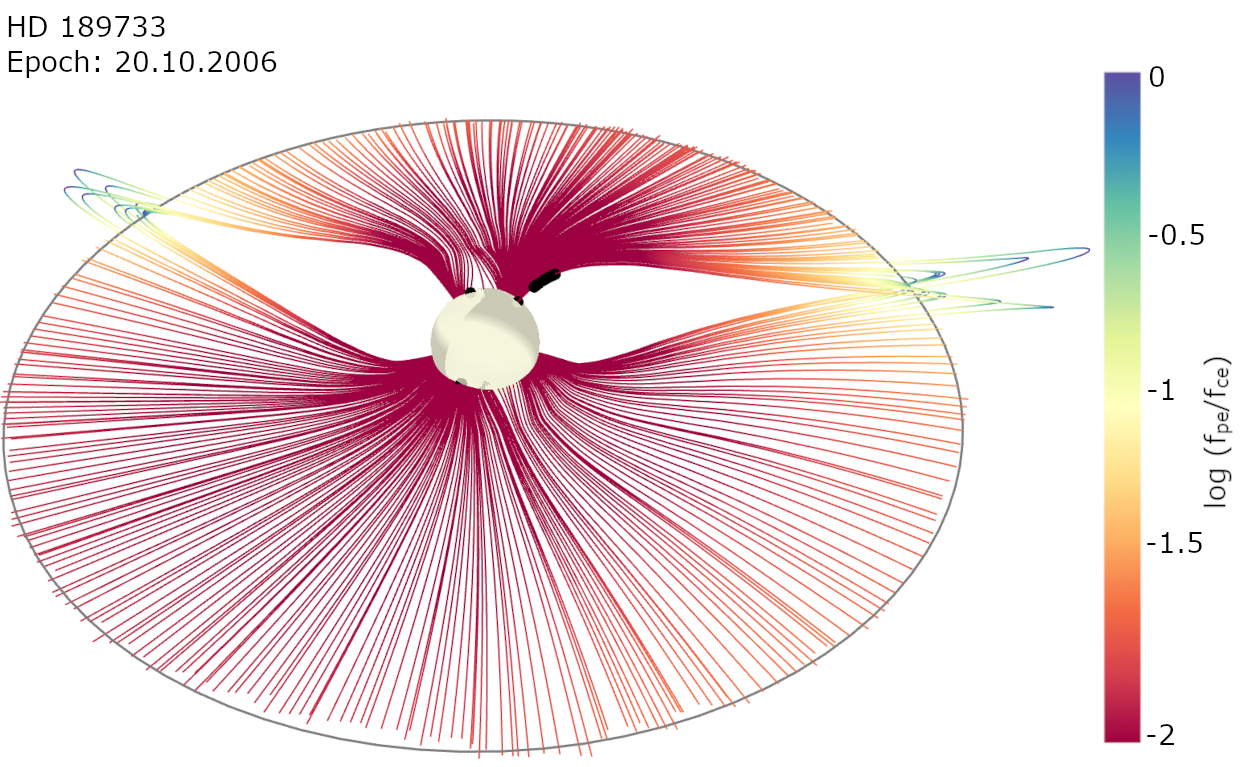}
    \caption{Same caption as Figure~\ref{fig:Tauboo-fpfc}, but for the case of HD~189733 at the epoch of 20 October 2006. The reader is invited to view the interactive version at \href{https://judy94.github.io/Star-system-FP-FC-visualization/HD189733_fpfce_source.html}{``HD~189733 system epoch animation''}.}
    \label{fig:HD189733-fpfc}
\end{figure}

% The explanation of the time-frequency pattern in Figure~\ref{fig:HD189733_frequency} follows the same reasoning as for HD~179949. 
Additionally, compared to HD~179949 and Tau~Boo, the plasma environment of HD~189733 is largely dominated by regions where the cyclotron frequency exceeds the plasma frequency, $f_{\rm ce} > f_{\rm pe}$ with \textbf{$\log \frac{f_{\rm ce}}{f_{\rm pe}} \sim -2$
}~(Figure~\ref{fig:HD189733-fpfc}).
As previously mentioned, propagation effects explain why only portions of the expected O-shaped emission pattern are visible and why the arcs appear stretched, distorted, or interrupted depending on the observer's latitude and the planet's orbital motion.
For example, the open parenthesis ``C-shape'' and closed parenthesis ``inverted C-shape'' of magnetic northern hemisphere emissions (red in Figure~\ref{fig:low_latitude_pattern}) indicate that the top part of the inverted U (yellow in Figure~\ref{fig:mid/high-latitude_pattern}) is not visible to the observer, either because high-frequency radio emission cannot escape the sources or the emission is reflected away from the observer's point of view.

Furthermore, we also take into account near-source refraction effects on the iso-$f_{\rm ce}$~(ie., waves emitted in the direction of increasing $B$). Emissions are permitted only when the wave frequency exceeds the cutoff frequency ($N > 0$), while waves approaching the cutoff~($N \rightarrow 0$) are absorbed or reflected. Consequently, the model does not assume that all generated emissions escape toward the observer.
Overall, these features bear a strong resemblance to arc-shaped radio emissions observed in planetary magnetospheres, such as the ones produced by the Jupiter-Io interaction, well modeled with ExPRES in \cite{Louis2017a} (``C'' and inverted-``C'' shapes, closed ``O'' or even more vertical structures).
Both the variations in this angle and the planet's orbital period can cause arcs to appear more stretched or distorted.

Since both HD~179949 and HD~189733 show predicted emission frequencies above the ionospheric cutoff in some cases, we explored how variations in stellar wind properties influence their radio emission. To quantify these variations, we use $\frac{S}{B}$, the stellar wind Poynting flux per unit magnetic field, which measures how effectively the plasma flow converts its motion into electromagnetic energy relative to the field strength. A higher $\frac{S}{B}$ indicates that, for the same $B$, the wind transports more electromagnetic energy (linked to a denser or faster flow).

Adjusting the stellar $\frac{S}{B}$ changes the mass-loss rate and wind density, which in turn shifts the AS and modifies the SPI environment~\citep{Hazra2021}. The range of $\frac{S}{B}$ for each star is estimated from physically motivated stellar mass-loss rates. These rates are derived from the observed X-ray surface fluxes using the empirical mass-loss--X-ray relation of \citet{Wood2021}, with a factor-of-two variation included to reflect typical observational uncertainties. Further details are provided in Appendix~\ref{app:stellarwind_conditions}.
Building on this, we explore variations of $\frac{S}{B}$~(same approach as previously described) for the full sample from \citet{Chebly2023}, investigating whether other stars with exoplanets can host planets orbiting within their AS; this analysis is presented in Appendix~\ref{app:Star systems from star sample}.

\section{Radio Power estimate}
\label{sec:radio power}
Exceeding Earth’s atmospheric cutoff is necessary but not sufficient for detection. The intensity of the radio must also be strong enough to surpass the sensitivity limits of current and upcoming ground-based telescopes. In this section, we focus on estimating the radio power, based on the Radio Magnetic Scaling Law \citep[or RMSL, ][]{Zarka2018, Zarka2025}, from each system and assessing their observability against telescope detection thresholds.

We use the 3D stellar wind simulations to extract the physical parameters needed to estimate the planetary radio emission. This will include extracting the wind total pressure\footnote{Sum of thermal, magnetic, and ram pressures} ($P_{\mathrm{tot}}$), magnetic field strength~($B_{\rm \perp}$), and stellar wind velocity ($\mathbf{V}_{\mathrm{wind}}$) at the planet orbit. These quantities allow us to calculate the magnetic power incident on the planetary magnetosphere~($P_{\rm mag}$, in Watts).% and the corresponding radio power.
\begin{equation}
    P_{\mathrm{mag}}~[\mathrm{W}] = \frac{B_{\rm \perp}^2}{\mu_0} V_{\mathrm{rel}} \pi R_{\mathrm{M}}^2
    \label{eq:Pmag}
\end{equation}

\noindent
where $B_{\rm \perp}$ is the total magnetic field strength at the orbital distance (in T), $\mu_0 = 4\pi \times 10^{-7}~\mathrm{H/m}$ is the vacuum permeability, $V_{\mathrm{rel}}$ is the relative velocity between the stellar wind and the planet (in m/s, see Equation~\ref{eq:vrel}), and $R_{\mathrm{M}}$ is the magnetospheric standoff radius (in meter, Equation~\ref{eq:RM}). The interaction cross-section is $\pi R_{\mathrm{M}}^2$.
\begin{equation}
    \mathbf{V}_{\mathrm{rel}} = \mathbf{V}_{\mathrm{wind}} - \mathbf{V}_{\mathrm{orbit}}
    \label{eq:vrel}
\end{equation}

\noindent
We assume a circular orbit, as such, the orbital velocity is given by $\mathbf{V}_{\mathrm{orbit}} = \sqrt{\frac{G M_{\star}}{r}}\, \boldsymbol{\hat{\varphi}}$, where $G$ is the gravitational constant, $M_{\star}$ is the stellar mass, and $r$ is the orbital radius. The unit vector $\boldsymbol{\hat{\varphi}}$ is azimuthal in spherical coordinates. 

Since our planets orbit within the stellar AS, the stellar wind is sub-Alf\'enic at their orbital distances. 
In this regime, there is no bow-shock on the star-planet direction \citep{Neubauer1980,Saur2013,Vidotto2015}, and the interaction is instead mediated by Alfv\'en wings \citep{Strugarek2015}. 
Therefore, the classical $R_{\mathrm{M}}$ distance should not be interpreted as a physical shock boundary. It serves as a proxy for the effective size of the planetary magnetosphere~(magnetopause radius) in the sub-Alfv\'enic configuration.
The magnetospheric standoff distance, is computed by balancing the planetary magnetic pressure with the total stellar wind pressure $P_{\rm tot}$ in watts and its given by~\citet{Gombosi2004, Shields2016}:
\begin{equation}
R_{\mathrm{M}}~[\rm m] = R_{\rm p} \left( \frac{B_{\rm p}^2}{2\mu_0} P_{\mathrm{tot}} \right)^{1/6}
\label{eq:RM}
\end{equation}
where $R_{\rm p}$ is the planetary radius and $B_{\rm p}$ is the equatorial magnetic field strength.

Since $B_{\rm p}$ at the boundary is not constrained for exoplanets, we explore a physically motivated range of values to assess their impact on the predicted radio emission.
For the lower limit, we assume a negligible planetary magnetic field, in which case the interaction with the stellar wind is governed purely by the planetary cross-section, $\pi R_{\rm p}^2$ which means $R_{\rm m}$ is simply approximated as $R_{\rm p}$. For the upper limit, we adopt $B_{\rm p} = 40~\mathrm{G}$, consistent with dynamo models for hot Jupiters under strong stellar irradiation~\citep{Yadav2017, Cauley2019}. While even stronger magnetic fields (up to $\sim$kG) are theoretically possible in very young planets \citep{Hori2021}, such extreme values are unlikely for the mature systems considered here.

The expected planetary radio power~($P_{\mathrm{radio}}$) is estimated using the empirical linear RMSL of the form $P_{\mathrm{radio}} = \epsilon \times P_{\mathrm{mag}}^{\alpha}$, where $\epsilon$ is an efficiency factor representing the fraction of magnetic energy converted into radio emission, and $\alpha$ is a power-law index~$\sim 1$~(\citealt{Zarka2018}).
The RMSL consistently accounts for both aurora magnetospheric and satellite-induced emissions at Jupiter, the latter being an analogue to SPI. Only the incident Poynting flux of the flow correlates with the emitted radio power; the apparent relation to the kinetic component is merely coincidental due to its similar heliocentric dependence and fails for satellite-induced emissions \citep[see Fig.~7a of][]{Zarka2018}. The most relevant references for the RMSL are \citet{Zarka2001}, \citet{Zarka2007}, \citet{Saur2013}, \citet{Zarka2018}, \citet{Callingham2024}, and \citet{Zarka2025}. In \citet{Zarka2018}, a linear fit ($\alpha=1$) yields a conversion efficiency $\epsilon \approx 2\times10^{-3}$, while an unconstrained fit gives $\alpha=1.15$; we adopt the linear case as the simplest and most robust. Further MHD simulations \citep{Varela2016,Varela2018,2022SpWea..2003164V,2024A&A...688A.138P} and analytical studies \citep{Nichols2016} suggest a more conservative range of $\epsilon \in [10^{-4}, 2\times10^{-3}]$.

Although the empirical scaling has been demonstrated predominantly for super-Alfv\'enic solar-wind coupling (\(M_A>1\); e.g. Earth and the outer planets), close-in star-planet systems often operate in the sub-Alfv\'enic regime (\(M_A<1\)). In this case, the appropriate Solar-System analogue is the Io/Ganymede--Jupiter interaction: there is no solar wind inside Jupiter’s magnetosphere, but the dense, corotating and slightly outward-drifting magnetospheric plasma constitutes an effective sub-Alfv\'enic flow in the satellite frame, generating stationary Alfv\'en wings that transport Poynting flux to the auroral source region (e.g.~\citealt{Saur2013,Zarka2018}).

It could be argued that planetary radio emissions are predominantly controlled by internal magnetospheric processes, implying no straightforward correlation with the incident solar wind power. However, extensive observations demonstrate a clear influence of the solar wind. For example, Saturn’s Kilometric Radiation (SKR) is strongly modulated by the solar wind \citep{Desch1982,Desch1983}, Earth’s Auroral Kilometric Radiation shows a direct correlation with solar wind parameters \citep{Gallagher1981}, and low-frequency Jovian emissions are similarly solar-wind controlled \citep{ZarkaGenova1983}. A comprehensive review is given in \citet{Zarka2001}. 
In the case of Jupiter, the broadband Kilometric Radiation is mostly modulated by solar wind variability~($\sim$ one quarter of the modulation \citealt{Zarka2001}), while the decametric~(DAM)\footnote{DAM = Decametric emission, at $\sim$10--40~MHz, the strongest Jovian auroral radio component.} and hectometric~(HOM)\footnote{HOM = Hectometric emission, at $\sim$0.3--3~MHz, part of Jupiter’s auroral spectrum.} emissions are predominantly driven by internal processes \citep{Zarka2021} related to the planet’s rapid rotation and its associated magnetospheric dynamics or satellite-magnetic field interactions~($\sim$ three quarters of the modulation). 

Nonetheless, these internal rotational effects are fundamentally coupled to the morphology of the magnetosphere and the location of the co-rotation breakdown boundary, typically occurring between 20 and 50~$R_{\rm J}$. Indeed, all auroral jovian radio components are sensitive and can be triggered by a strong magnetospheric compression by the solar wind \citep{Louis2023b}. Thus, the solar wind indirectly influences even the internally driven emissions by shaping the magnetospheric environment within which these processes operate.

% \citet{Zarka2018} propose two scaling laws: one links incident magnetic power to total planetary radio emission, and the other focuses on satellite–planet interactions (e.g., Io and Ganymede with Jupiter). Since our study targets SPI~(Satellite-Planet Interaction) radio emissions, we use the latter scaling law ($\epsilon = 2 \times 10^{-3}$, $\alpha = 1$), despite its limited data. This emission arises from the interaction of Jupiter’s corotating plasma with the ionosphere of Io or magnetosphere of Ganymede, rather than from the planet itself.

\begin{figure}[htbp]
  \centering
  \includegraphics[width=1\columnwidth]{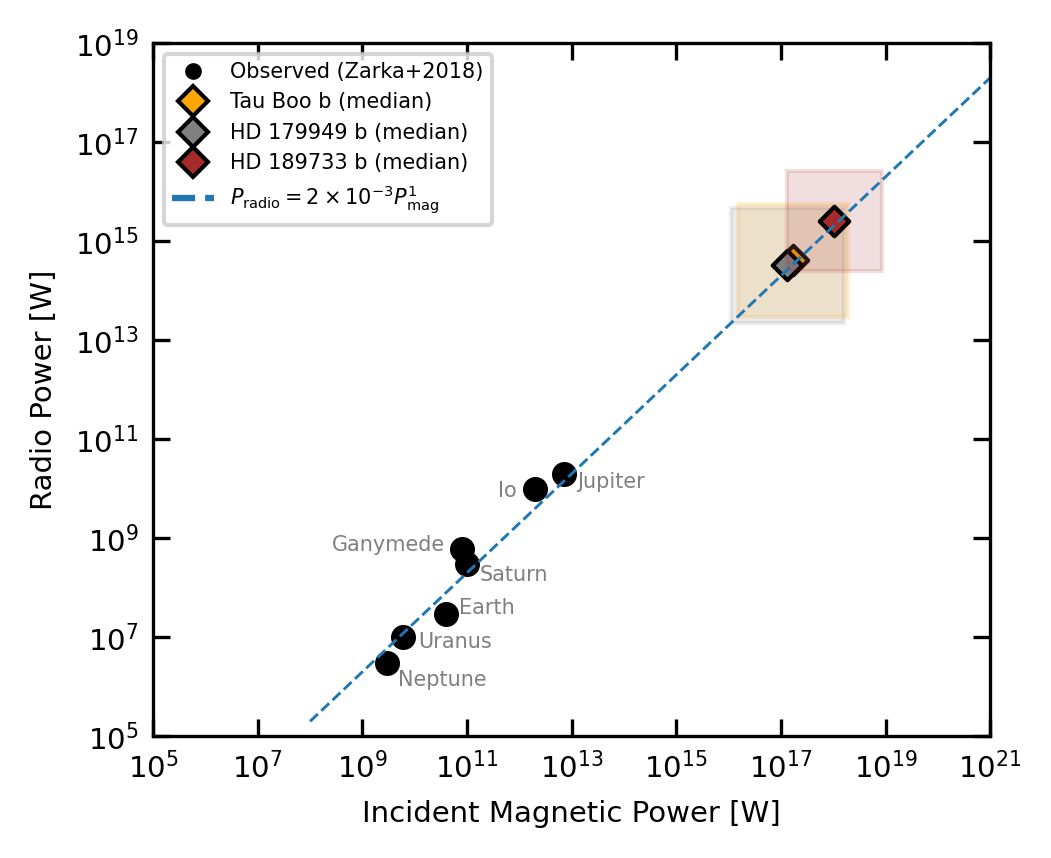} % or .png/.jpg
   \caption{
Radio Magnetic Scaling Law adapted from~\citet{Zarka2018} on which only the Satellite-Planet scaling law is drawn as the dashed blue line. The points corresponding to auroral planetary radio emissions are left for information. Predicted average radio power due to star--planet interaction for the three studied systems are added to the RMSL: Tau~Boo~(orange), HD~179949~(gray), and HD~189733~(red), compared with observed radio powers in the Solar System~(circle~marker). Colored boxes show model ranges obtained by varying the planetary magnetic field $B_{\rm p}$ from negligible to $40\,\mathrm{G}$; the box edges are the extrema on both axes. Diamonds mark the log–midpoint~(geometric center) of each box.}
  \label{fig:Radio_emission}
\end{figure}
\noindent

In Figure~\ref{fig:Radio_emission}, we compare the observed radio power from Solar System satellites and planets (black circles) to the predicted power from our target systems (diamond symbols). It is important to note that, for Jupiter, the ``Io'' point specifically represents the Io-induced emission. The ``Jupiter'' point, on the other hand, corresponds exclusively to the planet's non-satellite, auroral radio emission and does not include contributions from Io or Ganymede. The emitted radio power can be $10^{3}$ to $10^{6}$ times greater than what is observed from Jupiter, in agreement with the predictions of~\citet{Zarka2018}.
Interestingly, Tau~Boo is expected to produce radio power comparable to that of HD~179949, even though its emission frequency is lower. 

We now evaluate whether the SPI-induced radio power falls within the detection limits of current and upcoming ground-based radio telescopes, considering the full range of investigated planetary magnetic field strengths ($B_{\rm p}$). In particular, we focus on instruments whose frequency coverage overlaps with the simulated radio emission of our selected systems: the New Extension in Nançay Upgrading LOFAR (NenuFAR), the LOw-Frequency ARray (LOFAR), and the Square Kilometre Array Low (SKA-Low). The relevant sensitivity parameters for each telescope are detailed in Appendix~\ref{app:Telescope sensitivity}.
Telescope sensitivity is characterized by the ratio of effective collecting area to system temperature, $A_{\rm eff}/T_{\rm sys}$~[$m^{\rm 2}/K$], which can be converted into the System Equivalent Flux Density (SEFD), expressed in janskys~(Jy), using the relation in Equation~\ref{eq:SEFD}. When SEFD values are explicitly provided in the telescope documentation or technical handbook, we adopt those directly; otherwise, we calculate them using Equation~\ref{eq:SEFD}.
\begin{equation}
\mathrm{SEFD\,[Jy]} = \frac{2\,k_{\rm B}}{A_{\rm eff}/T_{\rm sys}} \times 10^{26}
\label{eq:SEFD}
\end{equation}
where $k_{\rm B}$ is the Boltzmann constant.

To estimate the detection threshold, we apply the standard radiometer equation (Eq.~\ref{eq:radiometer}) as defined by \citet{Thompson2017}, where the root-mean-square (RMS) noise $\sigma$ is given by:
\begin{equation}
\sigma\,[\mathrm{Jy}] = \frac{\mathrm{SEFD}}{\sqrt{n_p\Delta\nu t}}
\label{eq:radiometer}
\end{equation}
with $\mathrm{SEFD}$ representing the system equivalent flux density, $n_p$ the number of polarizations, $\Delta\nu$ the observing bandwidth (Hz), and $t$ the integration time (seconds).

The minimum detectable radio power ($P_{\rm radio}$) from a source at distance $d$ is given by
\begin{equation}
P_{\rm radio}\,[\mathrm{W}] = \Omega\, d^2 \, S_{\rm min}\,[\mathrm{W\,m^{-2}\,Hz^{-1}}] \, \Delta\nu \,[\mathrm{Hz}]
\label{eq:Pmin}
\end{equation}
where $\Omega$ is the solid angle into which the emission is beamed. In \cite{Zarka2004} they have shown that $\Omega$ typically takes values of $\sim 0.16$ sr for emission from a single flux tube (e.g., Io–Jupiter or star–planet interactions) and $\sim 1.6$ sr for emission from an extended auroral oval. 
The minimum detectable flux density $S_{\rm min}$ is defined as the product of the required signal-to-noise ratio (SNR) and the root-mean-square (RMS) noise $\sigma$:
\begin{equation}
S_{\rm min}\,[\mathrm{Jy}] = \mathrm{SNR} \times \sigma
\label{eq:Smin}
\end{equation}
Unless stated otherwise, we use these equations to estimate telescope detection thresholds, adopting a conservative signal-to-noise ratio of $\mathrm{SNR} = 5$, consistent with previous low-frequency radio studies \citep{Lazio2010, Zarka2015}. For each instrument, we compute both lower and upper limits on the detectable radio power, corresponding to the most optimistic and conservative thresholds within the telescope’s operational frequency range. Detailed values of the sensitivity parameters used for the telescopes along with the results, are provided in Appendix~\ref{app:Telescope sensitivity}.

\begin{figure}[htbp]
  \centering
  \includegraphics[width=1\columnwidth]{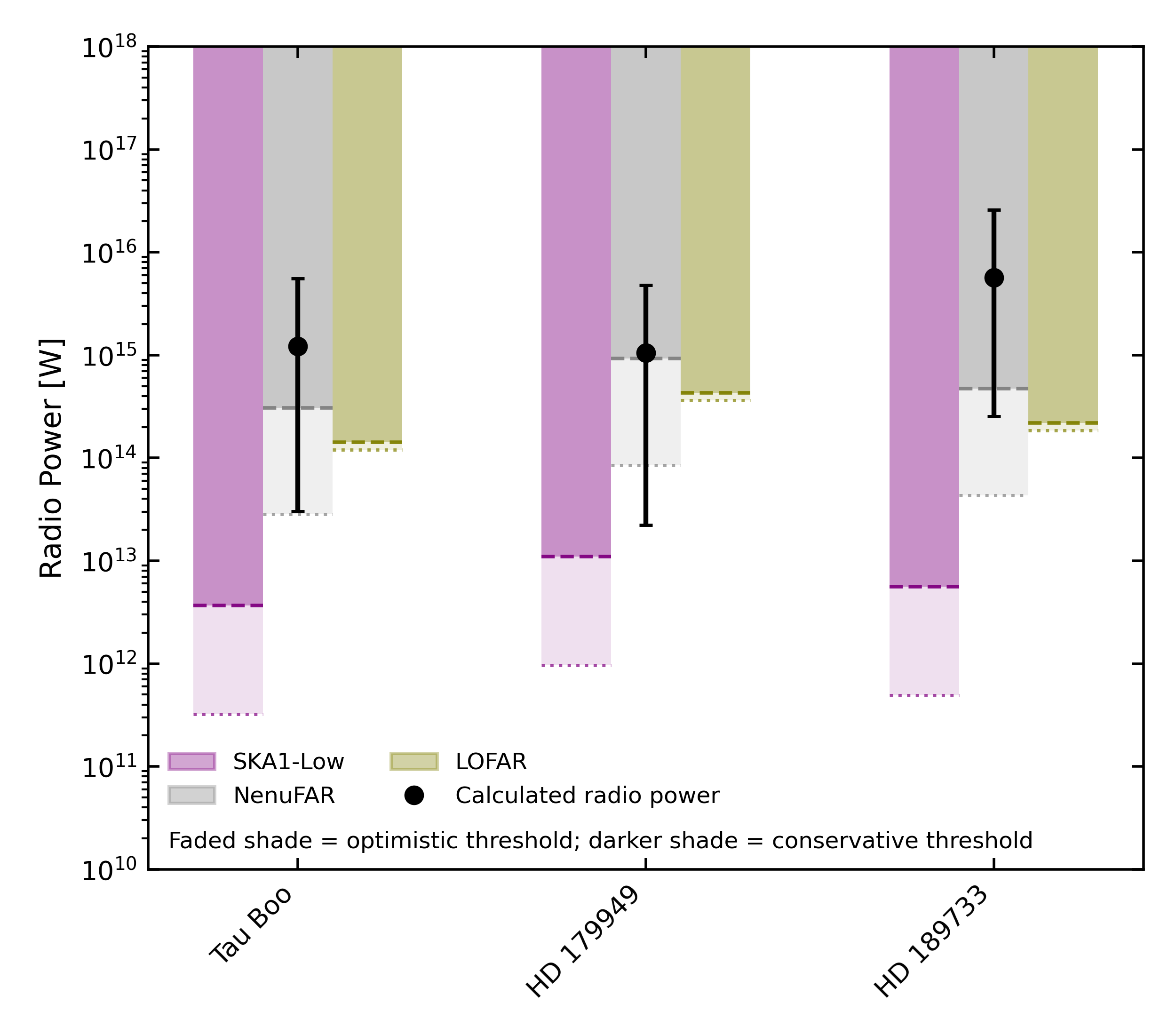} % or .png/.jpg
   \caption{Predicted radio power from star–planet interaction for Tau~Boo, HD~179949, and HD~189733 (black circles), with error bars representing the range of assumed planetary magnetic field strengths—from no field ($R_{\rm M} = R_{\rm p}$) to $B_{\rm p} = 40$~G. Colored bars indicate the minimum detectable radio power for SKA1-Low~(purple), NenuFAR~(gray), and LOFAR~(green) for each system. We show both conservative~(darker shade) and optimistic~(fainted shade) values of the telescopes based on the upper sensitivity limit. Dashed ticks mark the detection thresholds, they correspond to the bottom of each colored box. The values below that are undetectable by the respective instrument.}

  \label{fig:sensitivity}
\end{figure}

Figure~\ref{fig:sensitivity} displays the conservative~(dark shade) predicted planetary radio powers and the optimistic~(faded shades) minimum detection thresholds for each telescope. 
The detection thresholds for LOFAR (olive bar), NenuFAR (gray bar), and SKA1-Low (purple bar) are shown in both panels. 
NenuFAR and LOFAR exhibit similar sensitivities, while SKA1-Low provides the most favorable detection capability, allowing even weakly magnetized planets (down to $B_{\rm p} \sim 0.4$~G for HD~179949~b, Tau~Boo~b, and HD~189733~b) to be detected under conservative assumptions.
In the following analysis, we adopt the conservative thresholds, noting that using optimistic assumptions yields qualitatively similar results.

\begin{figure*}[hbt!]
    \centering
   \includegraphics[width=0.8\textwidth]{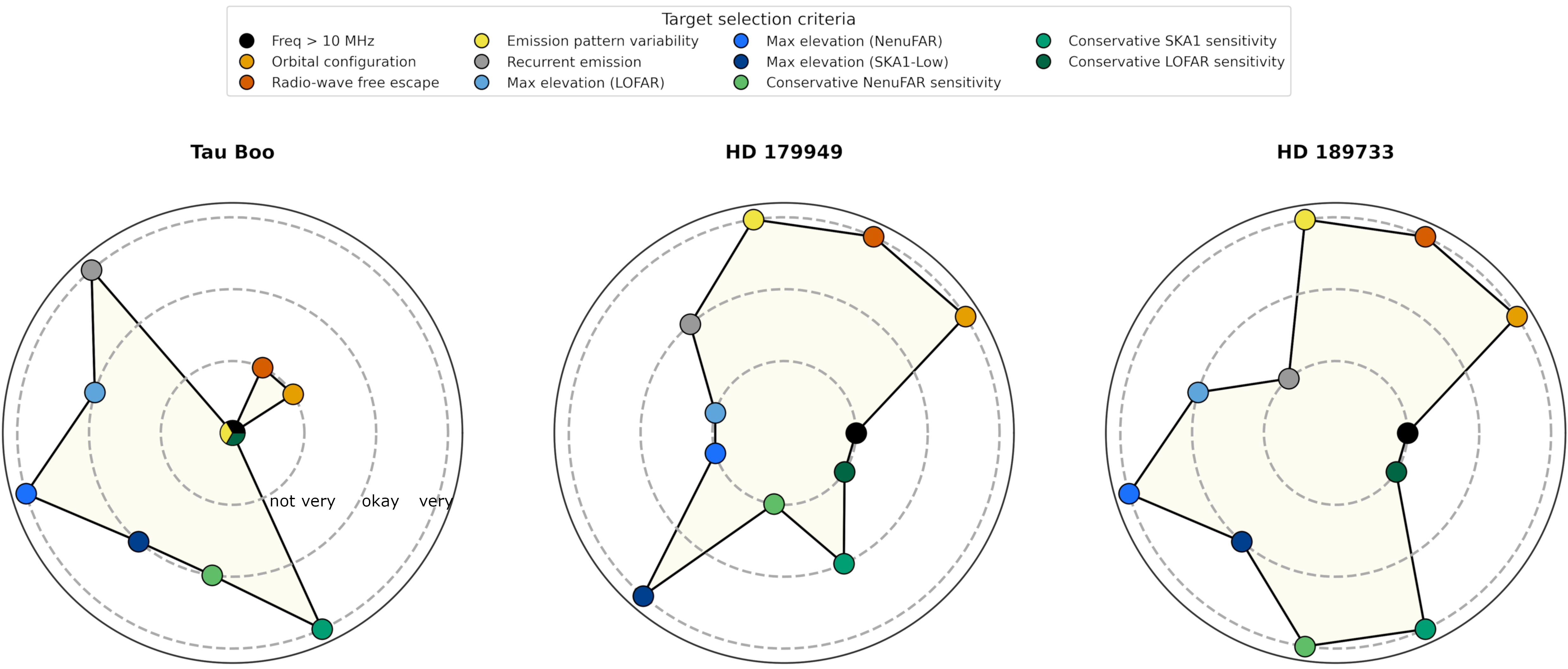}
        \caption{Radar plots summarizing the criteria used to identify the most favorable star–planet system for SPI-induced radio observations among Tau~Boo, HD~179949, and HD~189733 (from left to right). Each colored marker represents a specific criterion: detectable emission frequency above 10~MHz~(black), orbital configuration~(orange), radio-wave free escape~(red), emission pattern variablity~(yellow), recurrent emission (gray), maximum stellar elevation relative to each telescope (blue shades, light to dark: LOFAR, NenuFAR, SKA1-Low) and telescope sensitivity~(green shades, light to dark: NenuFAR, SKA1-Low, LOFAR). Four concentric dashed circles represent increasing levels of favorability: no, not very, okay, and very.}
    \label{fig:wrap_up}
\end{figure*}
\section{Target selection}
\label{target-selecton}

% After completing the analysis, we identify the most suitable targets for SPI-induced radio detection by evaluating each system across several key criteria. 
Following our analysis, we carry out a target-selection exercise, assessing each system against key criteria to identify the most promising candidates for SPI-induced radio detection. It should be noted that this evaluation is based on a single set of parameters; different assumptions or parameter choices could alter the ranking of targets. As such, the results are intended to guide target selection rather than provide definitive predictions, and a more comprehensive assessment would require ensemble simulations exploring a wider range of parameter combinations.

\subsection{Criteria}
Figure~\ref{fig:wrap_up} shows radar plots summarizing key observability criteria for Tau~Boo, HD~179949, and HD~189733 (left to right). The four concentric dashed circles represent a qualitative scoring scale from ``no'' to ``very''. This scoring is intended purely as a visual guide to facilitate comparison between systems and does not constitute an official classification.  
Colored symbols indicate individual criteria: frequency cutoff $<$10~MHz (black), orbital configuration (orange), radio-wave free escape (red), emission pattern variability (yellow), recurrent emission (gray), maximum stellar elevation relative to the telescope (blue shades: lighter to darker correspond to LOFAR, NenuFAR, SKA1-Low), and telescope sensitivity (green shades: lighter to darker correspond to NenuFAR, SKA1-Low, LOFAR).

%\titlespacing*{\subsubsection}{0pt}{2pt}{1pt} % {left}{before}{after}
\subsubsection*{Cut off frequency < 10~MHz}
We classify frequency cutoffs relative to the 10~MHz ionospheric limit: ``no'' for emission entirely below it (e.g., Tau~Boo), ``not very'' for emission just above 10~MHz (HD~179949, HD~189733), ``okay'' for emission extending beyond $\sim$100~MHz, and ``very'' for substantially higher frequencies.

\subsubsection*{Orbital configuration}
Orbital configuration of the exoplanetary system relative to the observer (see, for example,~\citealt{Lamy2023}): a magnetic pole-on orientation means the system is viewed nearly along its rotational axis, as if looking straight down onto the orbital plane like a spinning top. In this view, all visible sources are oriented away from the observer, which reduces direct visibility, as clearly illustrated in Figures~\ref{fig:mid/high-latitude_pattern} and \ref{fig:low_latitude_pattern}. In the case of Tau~Boo, the inclination of the system with respect to the observer results in more emission being directed away from the observer, making the configuration less favorable for detection, thus will be scored as ``not very''. In contrast, HD~179949 and HD~189733 have negligible inclination relative to the observer, leading to more favorable viewing conditions, thus scored as ``very''.

%\titlespacing*{\subsubsection}{0pt}{2pt}{1pt} % {left}{before}{after}
\subsubsection*{Radio-wave free escape}
Radio-wave free escape is limited by the plasma density surrounding the star, which can prevent emission from propagating out of the source region. This effect reduces the observable frequency range, even if the $B_{\rm p}$ is sufficient. As seen for Tau Boo, this results in a ``not very'' classification. For HD~179949 and HD~189733, the classification is ``very''~(Section~\ref{Prediction of radio emission}).

%\titlespacing*{\subsubsection}{0pt}{2pt}{1pt} % {left}{before}{after}
\subsubsection*{Emission pattern variability} 
Systems exhibiting diverse radio emission frequency patterns~(e.g., separated low- and high-frequency components, vertical or bracket-like structures) are preferred observational targets. Such diversity can indicate distinct magnetic geometries, emission regions, and beaming orientations, thereby increasing the likelihood that at least one emission component will be directed toward the observer during the observation window. This implies that, for the Tau~Boo system, the score will be ``no'' as there is only one pattern shape~(Figure~\ref{fig:tauboo_frequency} and Appendix~\ref{app:Entire Frequency-time diagram}). 
For HD~179949 and HD~189733, we observe different frequency-time patterns, so this will be scored as ``very''. 

%\titlespacing*{\subsubsection}{0pt}{2pt}{1pt} % {left}{before}{after}
\subsubsection*{Recurrent emission}
Recurrent emission is defined here as emission modulated by the orbital period of the planet. This implies that the emission will be visible more frequently, so that with an instrument subject to strong time constraints, the probability of detecting an event during any given observation is higher. For example, for Tau Boo this effect would improve the classification `very''; for HD~179949 ``okay''; and for HD~189733 from ``not very''.

\subsubsection*{Maximum elevation and observability}

The maximum elevation of a star~(the highest angle it reaches above the horizon when crossing the local meridian) determines its observability with a radio telescope. This value is independent of the observation date (see Appendix~\ref{app:max_elevation} for details). 

Based on the calculated maximum elevations (Table~\ref{tab:star_tel_coords_elev}) using equatorial-to-horizontal coordinate transformation, the observability of the target stars varies across the telescopes. 
We classify the observability of each target according to its maximum elevation ($\alpha_{\rm max}$). 
Targets with $\alpha_{\rm max} < 30^\circ$ are rated as ``not very'' observable, those with $30^\circ \le \alpha_{\rm max} \le 60^\circ$ as ``okay'', and those with $\alpha_{\rm max} > 60^\circ$ as ``very'' observable. 
This simple rating provides a qualitative measure of how favorably a target can be observed with each telescope.

For LOFAR, Tau~Boo and HD~189733 reach relatively high elevations of approximately 54.6$^\circ$ and 59.8$^\circ$, respectively, while HD~179949 only reaches 12.4$^\circ$.
For SKA1-Low, HD~179949 attains an elevation of 88$^\circ$, Tau~Boo reaches 45.8$^\circ$, and HD~189733 reaches 40.6$^\circ$. 
For NenuFAR, both Tau~Boo and HD~189733 are highly favorable with maximum elevations of 60.1$^\circ$ and 65.3$^\circ$ respectively, whereas HD~179949 remains poorly observable at 17.9$^\circ$. 
These ratings highlight the best target for each telescope: NenuFAR and LOFAR are most suitable for Tau~Boo and HD~189733, while SKA1-Low is ideal for HD~179949.
%\titlespacing*{\subsubsection}{0pt}{2pt}{1pt} % {left}{before}{after}
\subsubsection*{Telescope sensitivity}
A system is classified as ``very'' if the predicted radio power is detectable regardless of $B_{\rm p}$, and as ``no'' if it is undetectable for all $B_{\rm p}$. The label ``not very'' applies when detection is only possible for $B_{\rm p} \gtrsim 4.1\ \mathrm{G}$, while ``okay'' applies when detection requires a non-zero $B_{\rm p}$ but can occur for values both below and above 4.1~G, without spanning the full range. 
The scoring for each star–planet system is based on the results shown in Figure~\ref{fig:sensitivity}: for Tau~Boo, SKA1-low is rated ``very'', NenuFAR ``okay'', and LOFAR ``no''; for HD~179949, SKA1-low is ``okay'' under strict sensitivity, NenuFAR ``not very'', and LOFAR ``not very''; and for HD~189733, both SKA1-low and NenuFAR are ``very'', while LOFAR is ``not very''.

\subsection{Target assessment}
\label{sec:target_assessment}

The overall detectability of SPI-driven radio emission depends on both the intrinsic emission characteristics of the system and its observability from ground-based facilities. Our assessment combines physical emission criteria such as radio emission frequency domain relative to the ionospheric cutoff of Earth's atmosphere, plasma conditions, radio emission pattern variability, and recurrence. We also take into account instrumental constraints such as telescope sensitivity and source elevation with respect to the telescope altitude. The resulting classification reflects how these factors jointly influence the target selection. We note, however, that a more robust assessment would require ensemble simulations exploring a wider range of parameter combinations.

HD~179949 exhibits recurrent radio emission linked to its orbital period and a favorable, nearly pole-on geometry. This configuration increases the temporal coverage of emission peaks and enhances the probability of periodic signal detection. The system reaches a maximum elevation of $88^{\circ}$ at SKA1-Low, enabling near-zenith observations and optimal sensitivity, while it remains low on the horizon from northern sites ($17.9^{\circ}$ for NenuFAR and $12.4^{\circ}$ for LOFAR). Although its detectability decreases under conservative sensitivity assumptions, its high elevation and emission recurrence make it particularly suited for SKA1-Low monitoring campaigns in the southern hemisphere.

HD~189733 generates the highest predicted radio powers among the considered systems. Its emission lies systematically above the ionospheric cutoff and escapes efficiently through the stellar wind plasma, ensuring robust detectability across a wide range of planetary magnetic field strengths. With maximum elevations of $65.3^{\circ}$, $59.8^{\circ}$, and $40.6^{\circ}$ for NenuFAR, LOFAR, and SKA1-Low, respectively, the system is observable from both hemispheres with sufficient altitude for high signal stability. Its emission pattern variability further broadens the range of favorable viewing geometries, making HD~189733 the most detectable target overall. However, observing this system will be costly, as the radio emission is not as frequent as for the other systems. 

Tau~Boo is characterized by strong stellar magnetism and frequent magnetic polarity reversals that influence the emission geometry and field topology. 
However, its dense stellar wind limits radio-wave escape, confining the emission below the ionospheric cutoff and strongly reducing detectability. 
Despite relatively high visibility from northern sites (60.1$^\circ$ at NenuFAR and 54.6$^\circ$ at LOFAR), emission suppression in the high-density wind regime makes ground-based detection unlikely. 
Nevertheless, its short magnetic cycle ($\sim$120~days) makes it a valuable laboratory for studying the temporal evolution of SPI emission across magnetic reversals.
Similar cycle-driven variability might take place on the other two stars, though on different timescales: HD~179949 has shown evidence of polarity reversals on year-long timescales \citep[e.g.][]{Fares2012}, while HD~189733 exhibited a polarity switch between 2006 and 2008 \citep{Fares2010}, indicative of a short magnetic cycle.

In conclusion, the preferred target depends on the observing
facility and scientific objective. HD~179949 offers the most favorable conditions for southern arrays such as SKA1-Low, combining recurrent emission and near-zenith visibility. HD~189733 provides the strongest and most persistent emission across all
configurations, maximizing the overall likelihood of detection.
Tau Boo, while less favorable observationally, remains 
relevant for probing SPI variability under rapidly evolving magnetic conditions.

\section{Caveats}
\label{sec:caveats}
Simulation tools have inherent limitations that stem from both model assumptions and the quality of input data.
ZDI is limited by line-of-sight integration, capturing only net circular polarization~(Stokes V). This causes cancellation of opposite magnetic polarities with comparable magnetic field strength~\citep{Lang2014, Kochukhov2020}, making ZDI insensitive to small-scale magnetic fields and underestimating total surface flux—especially in slow rotators with weak Doppler broadening. This underestimation propagates into stellar wind models (e.g.,~AWSoM, WindPredict-AW) that use ZDI maps as boundary conditions, leading to weaker surface fields, underestimated wind strength, mass-loss rates, and AS sizes.

This being said, we would like to highlight that the predicted emission frequency in our simulations follows directly from the field strengths provided by the ZDI maps. 
As such, there is little flexibility beyond the intrinsic precision of the maps themselves, which is difficult to quantify but sets the main limit on frequency accuracy. 
In contrast, the shapes of the time--frequency patterns we obtain are highly robust. 
As illustrated by the extended plots provided in the Appendix~\citep[see Appendix \ref{app:stellarwind_conditions}, and the Online Supplementary Material,][]{Chebly2025_SI}, these features persist across the different assumptions explored, and remain stable even when model parameters are varied. 
We therefore emphasize that while absolute frequency values are tied to the precision of the input maps, the morphological features of the predicted emission are a robust outcome of our approach.

In addition, our simulations do not follow the star's magnetic cycle; we use a single-epoch ZDI map and do not account for temporal variations in the magnetic field. While this approach allows us to explore the range of star–planet–observer configurations and the resulting radio emission, one should keep in mind that the star’s activity cycle can significantly affect the magnetic field strength and topology over time. Future studies should consider the impact of stellar activity cycles to capture more realistic temporal variations in the SPI and emission properties.

Moreover, ZDI’s effectiveness also depends on stellar rotation: fast rotators (e.g., active M dwarfs) require high temporal resolution to track rapidly evolving surface features, while slow rotators like Proxima Centauri pose challenges due to weak Doppler signals and long periods. Near-infrared instruments (SPIRou, CRIRES+) improve access to M~dwarfs but face issues such as molecular line blending and uncertain Land\'e factors~\citep{Cristofari2022,Hahlin2023}.

Stellar wind models for cool stars beyond the Sun lack direct observational constraints on coronal base densities and temperatures, necessitating reliance on empirical or theoretical scaling laws that introduce uncertainties. These limitations affect radio emission modeling tools (e.g., ExPRES), which depend on accurate magnetic field and density topologies to simulate radio emission frequencies. 

Additionally, ExPRES does not compute radio emission intensity; rather, it predicts the occurrence and visibility of radio sources under specified plasma and magnetic field conditions (as discussed previously in section~\ref{subsec:radio emission simulation}). It is worthnoting other model limitations that must be considered.
First, concerning the wave propagation, ExPRES includes refractive effects only locally, within or near the source region (see Section~\ref{subsec:radio emission simulation}), and does not account for large-scale propagation phenomena such as refraction through extended plasma structures, which can significantly alter the visibility of low-frequency radio emission~\citep{Louis2019}.
Second, the simulated time--frequency morphology is sensitive to uncertainties in several key system parameters: the topology of the large-scale stellar magnetic field, the spatial distribution of plasma density, the planetary orbital configuration~(which is a known parameter), and the energy distribution of the emitting electrons. When these parameters are well constrained, ExPRES can reliably model the beaming geometry of the emission and, in favorable cases, provide insight into local plasma conditions such as the minimum electron energy required for cyclotron emission~\citep{Hue2023}.

Overall, accurate simulation of stellar radio emissions in exoplanetary systems is challenged by several sources of uncertainty. First, the limited resolution and accuracy of stellar magnetic field maps from ZDI, which serve as simulation boundary conditions, introduce inherent limitations. Second, uncertainties in stellar wind models and coronal parameters further complicate the predictions. Despite these challenges, current modeling approaches remain the most reliable tools available for studying such complex star–planet interactions.

\section{Conclusions}
\label{sec:conclusion}  

This proof-of-concept study demonstrates that combining predicted radio emission frequencies from the ExPRES code with planetary radio power estimates from the Radio–Magnetic Scaling Law, based on 3D~MHD stellar wind simulations driven by ZDI magnetic maps, provides a robust framework for prioritising SPI radio targets. The method enables (i)~verification of whether SPI emissions occur at frequencies accessible to ground-based facilities, (ii)~prediction of characteristic structures in the emission periodograms, and (iii)~evaluation of whether the expected radio power is detectable by instruments such as LOFAR, NenuFAR, or SKA.

Applying this framework to Tau~Boo, HD~179949, and HD~189733 reveals that the optimal target depends on both the observing facility and the underlying scientific goal, as each system presents distinct advantages and limitations. Among them, HD 189733 appears to produce some of the strongest and most persistent emission across all configurations, suggesting a higher likelihood of detection.
HD~179949 offers the most favorable conditions for SKA1-Low, combining recurrent emission with near-zenith visibility, making it particularly well suited for dedicated monitoring campaigns. In contrast, Tau~Boo, though less favorable from an observational standpoint, remains scientifically valuable as a natural laboratory for studying the temporal variability of SPI-induced radio emission under rapidly evolving magnetic conditions.

We also show that SPI-induced radio emissions can exhibit a wide variety of temporal patterns~(not restricted to O-shaped profiles) arising from differences in system geometry, magnetic topology, and stellar wind variability. Our model provides predictions for both the timing and morphology of the expected signals. 
However, these results are sensitive to the choice of ZDI map used to drive the stellar wind model. 
Variations in the magnetic field reconstruction can lead to changes in the resulting wind solution and thus in the predicted shapes, timings, and fluxes of the radio signals. 
Therefore, the robustness of these predictions should be interpreted with caution, while still enabling direct testing against future observations to confirm or refute their SPI origin.

Although our analysis focuses on Tau Boo, HD~179949, and HD~189733, the methodology is applicable to any system with available ZDI maps and wind models. This framework provides a robust basis for prioritising targets and optimizing telescope scheduling for future SPI observations. Future work should incorporate stellar magnetic cycle variability, as changes in field topology can significantly affect SPI radio emission characteristics. The approach can also account for the impact of $S/B$ variations~(driven by changes in coronal base density and temperature) on the planet’s position relative to the AS, enabling the identification of additional promising targets. Tailored wind simulations for each star remain essential, as initial wind conditions strongly influence predicted emission frequencies. With continued advances in magnetic field reconstruction and stellar wind modelling, this method offers a scalable pathway for detecting and characterising magnetic SPI across a broader stellar sample, ultimately providing deeper insights into the magnetic environments of exoplanets.

\section*{Data Availability Statement}

The ExPRES radio emission simulations (configuration files, precomputed magnetic field lines, and results) presented in this paper have been gathered into a collection available at \url{https://doi.org/10.25935/4pks-d207} \citep{Chebly2025_SI}.They were computed using ExPRES Version 1.4.0 \citep{Louis2025}.
Owing to the large volume of the 3D stellar wind simulations, the complete datasets will be provided to the community upon reasonable request. 
Extracts of specific quantities discussed in this work are available from the corresponding author.

\begin{acknowledgements}
We thank the referee for constructive comments that helped tightening the presentation of this work. J.~J.~C. and A.S. acknowledge funding from the European Research Council project ExoMagnets (grant agreement no. 101125367) and the PLATO/CNES grant at CEA/IRFU/DAp. C.~L. and P.~Z. acknowledge funding from the European Research Council under the European Union’s Horizon 2020 research and innovation program (grant agreement no 101020459 - Exoradio).
J.~J.~C. also thanks Mr.~Jad Estephan for his assistance in executing the artistic illustration rendering using Adobe Illustrator.
\end{acknowledgements}

\appendix

\section{\mbox{Coupling stellar wind model to ExPRES}}
\label{app:coupling wind to ExPRES}

The MHD properties—magnetic field components and density—along individual field lines are stored in CSV files, but only those connecting the star to the planet are considered. Each file corresponds to a specific longitude and must follow the naming format "XX\_polarity\_LineNumber", where "XX" can be any identifying label. The polarity is indicated by "\_m\_" for magnetic field lines directed from the planet to the star (magnetic northern hemisphere) and "\_p\_" for lines from the star to the planet (magnetic southern hemisphere). The "\_Line000" segment contains the longitude index; for example, if there are 300 "\_m\_" lines, filenames range from "\_Line000" to "\_Line300". 

Each file begins with two header lines starting with "\#": the first specifies whether the field line is connected to the star (\#Field line is connected to star: True/False), and the second lists the columns as \#X [Rs], Y [Rs], Z [Rs], BX [G], BY [G], BZ [G], Rho [g/$\rm cm^{3}$]. 

The files in an archive (MFL\_B\_and\_density.zip) can be downloaded from the following addresses: \url{https://doi.org/0.25935/9c80-
w202} \citep[for Tau Bootis,][]{Chebly2025_ExPRES_TauBoo}, \url{https://doi.org/10.25935/ce9q-
n094} \citep[for HD 189733,][]{Chebly2025_ExPRES_HD189733}, \url{https://doi.org/10.25935/d1h2-z977} \citep[for HD 179949,][]{Chebly2025_ExPRES_HD179949}.

\titlespacing*{\section}{0pt}{6pt}{1pt} % 
\section{Different electron energy exploration}
\label{app:Entire Frequency-time diagram}
 
We investigate how changes in electron energy affect the radio emission patterns for a loss-cone distribution.  
We present results for Tau~Boo, HD~179949, and HD~189733~in the Online Supplementary Materials \citep[see][Figures~B1, B2, B3, respectively]{Chebly2025_SI},
%\ref{fig:tauboo_all_epochs_cases},\ref{fig:HD179949_all_epochs_cases}, \ref{fig:HD189733_all_epochs_cases} respectively)
showing seven epochs of simulated radio emission for different electron energies.
For context, we simulate electron energies of 1~keV, 20~keV, and 100~keV to provide a complete picture. This choice is guided by observations at Jupiter, where electron populations range from a few tens of eV to hundreds of keV, but where radio emissions are produced mainly by electrons with energies below $\sim$30~keV.  
Therefore, the 1~keV and 20~keV cases are the most relevant for interpretation.

As shown in previous studies \citep[e.g.,][]{Hess2008a, Louis2019, Zarka2025}, higher electron energies correspond to smaller beaming angles, resulting in less visible emission in both time and frequency.  
This also modifies the arc morphology due to changes in the beaming pattern, for example, as clearly seen in the Tau~Boo case~\citep[see][Figure~B1, first column]{Chebly2025_SI}, where both "U"-shaped and inverted "U"-shaped patterns emerge and the radio emission is more frequently visible. 
% Also in HD~179949~\citep[see][Figure~B2, last row, just before 07-04-10]{Chebly2025_SI}, the red emission evolves from a "C"-shaped structure~(1~kev) into a closed loop~(20~kev). 
Same goes for HD~189733~\citep[see][Figure~B3]{Chebly2025_SI}. 

\section{Different orbital phase}
\label{app:Ephemeris}

\begin{figure}[hbt!] 
     \centering
      \includegraphics[width=0.75\columnwidth]{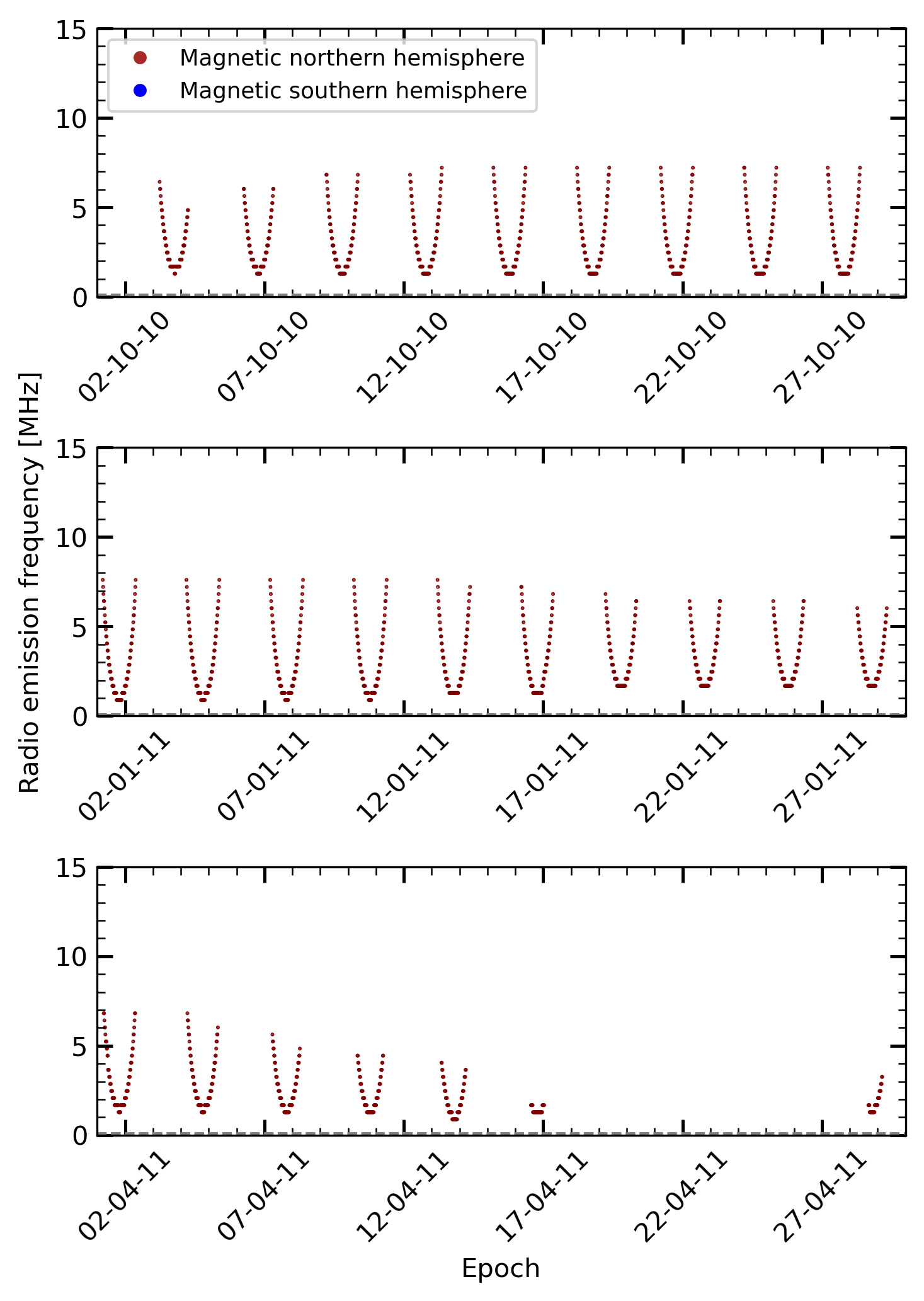}
      \caption{Same caption as Fig.~\ref{fig:tauboo_frequency}, but in this case we simulated the orbital phase at the beginning of the simulation in ExPRES at $\phi =2.5$ instead of $\phi =0$ which gave exactly the same results.}
      
    \label{fig:tauboo_frequency_diff_orbitalphase}
\end{figure}

To evaluate how the choice of ephemeris influences the predicted radio emission frequencies, we carried out a test on the Tau~Boo system. Since the effect is expected to be similar for the other systems considered in this study, we selected Tau~Boo as a representative case. Using the ExPRES code, we applied the orbital ephemeris and simulated the resulting radio emission for the case of a loss-cone electron distribution with an energy of 20~keV. 
In this test, we adopted a different orbital phase at the start of the ExPRES simulation, $\phi=2.5$ instead of $\phi=0$~(the orbital phase in the initial simulations done in Section~\ref{Prediction of radio emission}). With this modification, we obtained the results shown in Fig.~\ref{fig:tauboo_frequency_diff_orbitalphase}.

The comparison between simulations initialized at different orbital phases shows that the resulting radio emission frequencies and patterns are essentially unchanged regardless of the planet’s position at the start of the simulation. This indicates that the emission is primarily controlled by the large-scale star–planet magnetic interaction rather than by the exact orbital phase. In the context of observations, this means that uncertainties in the orbital ephemeris (e.g., in $T_0$ or $P$) may shift the timing of the signal within the orbital cycle but do not affect its range and morphology. Consequently, our predictions remain robust for observational planning, as the choice of ephemeris does not critically bias the expected radio signature.

\titlespacing*{\section}{0pt}{6pt}{1pt} % 
\section{Poynting flux exploration}
\label{app:stellarwind_conditions}

The plasma conditions of the stellar wind near the star strongly influence the magnetic interaction between the star and its planet. By varying these conditions, particularly through changes in $\frac{S}{B}$, we can examine how the resulting SPI affects the radio emission properties. As shown by \citet{Hazra2021}, $\frac{S}{B}$ is a key driver of mass loss in magnetically active stars, making it a useful proxy for our modeling.

In \citet{Chebly2023}, the $\frac{S}{B}$ ratio was fixed at a canonical value typical of the Sun, $1.1 \times 10^6~\mathrm{J\,m^{-2}\,s^{-1}\,T^{-1}}$. Here, we extend this approach to the exoplanet host stars HD~189733 and HD~179949 by estimating $\frac{S}{B}$ from a physically motivated range of stellar mass-loss rates ($\dot{M}$). To remain consistent with observational constraints, we derive $\dot{M}$ from the observed X-ray surface fluxes ($F_{\rm X}$) using the empirical $\dot{M}$–$F_{\rm X}$ relation of \citet{Wood2021}, adopting a factor-of-two variation to reflect typical observational uncertainties. While this method does not capture detailed magnetic topologies, it provides a realistic first-order approximation of the wind properties. We then infer $\frac{S}{B}$ corresponding to these $\dot{M}$ values using the near-linear $\dot{M}$–$\frac{S}{B}$ relation established by \citet{Hazra2021} under solar maximum conditions (CR~1975). The stellar properties and resulting $\frac{S}{B}$ ranges are summarized in Table~\ref{tab:star_info}. The latter also lists the $\overline{AS}_{\rm eq}$ obtained from wind simulations with $\frac{S}{B} = 1.1 \times 10^6~\mathrm{J\,m^{-2}\,s^{-1}\,T^{-1}}$ from \citet{Chebly2023}.

 % For systems without direct X-ray measurements, we estimate the stellar X-ray luminosities ($L_{\rm X}$) from their rotation periods and masses using the empirical activity–rotation relation of \citet{Pizzolato2003}. 
\begin{table*}[hbt!]
\centering
\caption{
List of stellar parameters for the 21 stars in our sample, along with their associated exoplanets where applicable.
}
\resizebox{0.6\textwidth}{!}{%
\begin{tabular}{l c c l c c c c c c}
\hline
Star name & Radius ($R_\odot$) & Spectral type & Exoplanet & $a$~(AU) & $a$~($R_\star$) & $\overline{AS}_{\rm eq}$~(AU) & $\overline{AS}_{\rm eq}$($R_\bigstar$) & $\log{L_{\rm X}}$ & $\frac{S}{B}_{\rm estimated}$~\,($\mathrm{J m^{-2} s^{-1} T^{-1}}$)\\
\hline
Tau~Boo         & 1.46 & F7V & tau~boo~b      & 0.04869 & 7.170        & 0.0670  & 9.99  & 28.99 & $7.68\times 10^5$ - $4.25\times 10^6$  \\ 
HD~179949       & 1.19 & F8V & HD~179949~b    & 0.04439 & 8.020        & 0.0567  & 10.24 & 28.30\textsuperscript{a} &  $1.06\times 10^5$ - $5.87\times 10^5$ \\ 
HD~35296        & 1.10 & F8V & --             & --      &       --       & 0.2069  & 40.43 & 29.33 &-- \\ 
HN~Peg          & 1.04 & G0V & HN~peg~b       & 773     & 159802.884   & 0.0660  & 13.65 & 29.19 & $1.06\times 10^4$ - $5.92\times 10^4$\\ 
HD~190771       & 1.01 & G2V & --             & --      &       --       & 0.0580  & 12.35 & 29.13 & -- \\ 
TYC~1987-509-1  & 0.83 & G7V & --             & --      &     --         & 0.0597  & 15.47 & 27.3\textsuperscript{b} &-- \\ 
HD~73256        & 0.89 & G8V & HD~73256~c     & 3.8     & 917.977      & 0.0444  & 10.73 & 27.7\textsuperscript{b} & $3.44\times 10^4$ - $1.91\times 10^5$ \\ 
HD~130322       & 0.83 & K0V & HD~130322~b    & 0.0929  & 24.0644      & 0.0313  & 8.12  & 26\textsuperscript{b} & $7.98\times 10^2$ - $4.42\times 10^3$\\ 
HD~6569         & 0.76 & K1V & --             & --      &      --        & 0.0574  & 16.25 & 27.4\textsuperscript{b} & --\\ 
$\epsilon$~Eri     & 0.72 & K2V & Epsilon~eri~b  & 3.5     &      1045.13        & 0.0456  & 13.64 & 28.33 & $1.2\times 10^5$ - $6.71\times 10^5$ \\ 
HD~189733       & 0.76 & K2V & HD~189733~b    & 0.03126 & 8.84         & 0.0474  & 13.40 & 28.46 & $1.66\times 10^5$ - $9.21\times 10^5$ \\ 
HD~219134       & 0.78 & K3V & HD~219134~b    & 0.03838 & 10.579       & 0.0302  & 8.33  & 26.85\textsuperscript{c} & $5.26\times 10^3$ - $2.91\times 10^4$\\ 
                &      &     & HD~219134~c    & 0.06466 & 17.822       &     --    &  --     &   --    & \\ 
                &    &    & HD~219134~d    & 0.2345  & 64.637       &    --     &  --     &   --    & \\ 
                &    &    & HD~219134~f    & 0.1453  & 40.0506      &   --      &  --     & --      & \\        
                &    &    & HD~219134~g    & 0.3753  & 103.448      &   --      &  --     &  --     & \\ 
                &     &     & HD~219134~h    & 2.968   & 818.1025     &  --       &   --    &    --   & \\ 
TYC~6878-0195-1 & 0.64 & K4V & --             & --      &       --       & 0.0689  & 23.16 & 26.88\textsuperscript{b} &-- \\ 
61~Cyg~A        & 0.62 & K5V & --             & --      &       --       & 0.0344  & 11.93 & 27.57 & \\ 
HIP~12545       & 0.57 & K6V & --             & --      &     --         & 0.0648  & 24.46 & 27.8\textsuperscript{b} & --\\ 
TYC~6349-0200-1 & 0.54 & K7V & --             & --      &      --        & 0.0527  & 21.00 & 28.2\textsuperscript{b} & --\\ 
DT~Vir          & 0.53 & M0V & Ross~458~c     & 1100    & 446226.415   & 0.0721  & 29.23 & 29.16 & $6.26\times 10^5$ - $3.47\times 10^5$\\ 
GJ~205          & 0.55 & M1.5V & --           & --      &     --         & 0.0407  & 15.91 & 27.66 & --\\ 
EV~Lac          & 0.30 & M3.5V & --           & --      &   --           & 0.1663  & 119.16 & 28.99 & --\\ 
YZ~Cmi          & 0.29 & M4.5V & --           & --      &      --        & 0.1015  & 75.23 & 28.57 & --\\ 
GJ~1245~B       & 0.14 & M6V  & --            & --      &      --        & 0.0248  & 38.12 & 27.47 & --\\ 
\hline
\end{tabular}%
}
\vspace{0.2em}
\tablefoot{The table lists, for each star: its name, radius in solar units ($R_\bigstar$), spectral type, and associated exoplanet (if any; a dash indicates no confirmed planet). The planetary semi-major axis $a$ is given both in AU and in units of stellar radii ($R_\odot$). The average equatorial Alfv\'en surface radius ($\overline{AS}_{\rm eq}$) is reported in AU and in stellar radii. The following columns show the stellar X-ray surface luminosity ($\log L_{\rm X}$) and the $\frac{S}{B}_{\rm estimated}$ range estimated from scaling laws; we only consider $\frac{S}{B}_{\rm estimated}$ ranges for stars hosting at least one planet. All stellar parameters are taken from \citet{Chebly2023} and references therein, while X-ray luminosities are from the ROSAT All-Sky Survey unless noted otherwise. \textsuperscript{a}~From~\citet{Acharya2023}. \textsuperscript{b}~Estimated using the empirical relation of~\citet{Pizzolato2003}. \textsuperscript{c}~From~\citet{Schmitt2004}.
} \\

\label{tab:star_info}
\end{table*}

\begin{figure*}[hbt!] 
     \centering
      \includegraphics[width=0.5\textwidth]{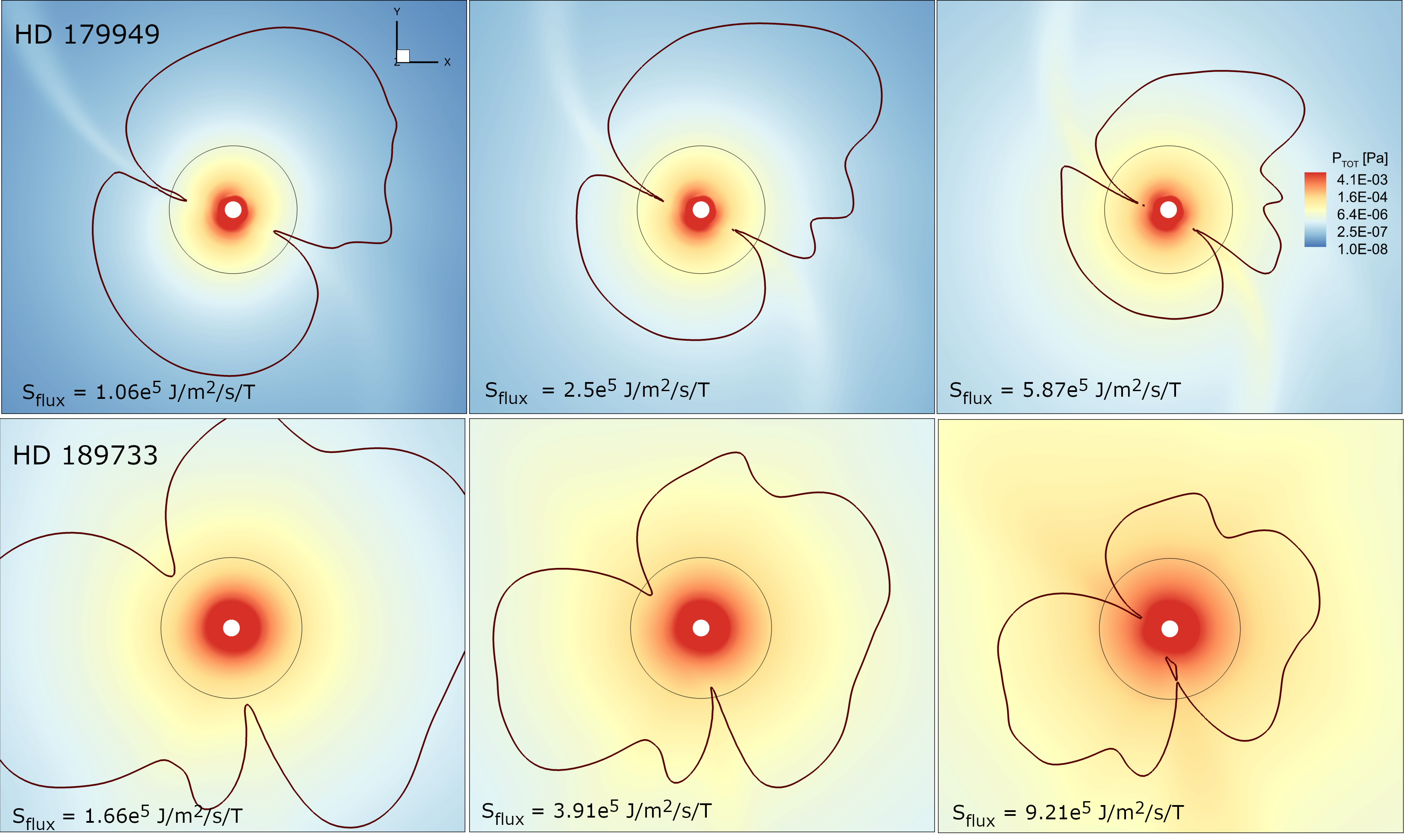}
      \caption{Different Poynting flux ($\frac{S}{B}$) scenarios for HD~179949 (top) and HD~189733 (bottom) illustrating how changes in stellar wind influences star--planet interaction-induced radio emission. The colormap shows the total stellar wind pressure (sum of thermal, ram, and magnetic pressures) in pascals, with orange-red regions indicating higher pressure. The solid black circle marks the planetary orbit, and the brown contour outlines the shape of the Alfv\'en surface at the equator. HD~189733 exhibits a higher total pressure than HD~179949.}
      \label{fig:Sflux_eploration}
\end{figure*}

An increase in $\frac{S}{B}$ enhances the energy available for Alfv\'enic perturbations, pushing the AS inward. For a given global magnetic field, higher $\frac{S}{B}$ strengthens wind driving, resulting in faster acceleration and a smaller AS. This effect is illustrated in Figure~\ref{fig:Sflux_eploration}, with the upper row showing HD~179949 and the lower row HD~189733 for different $\frac{S}{B}$ values. In all cases studied, the planet’s orbit remains almost entirely within the AS regardless of $\frac{S}{B}$. For HD~179949, increasing $\frac{S}{B}$ from $1.06\times 10^5$ to $5.87\times 10^5~\mathrm{J\,m^{-2}\,s^{-1}\,T^{-1}}$ (a factor of $\sim$5) reduces $\overline{AS}_{\rm eq}$ from $\sim$18 to $\sim$14\,$R_\bigstar$ (28.6\%). For HD~189733, increasing $\frac{S}{B}$ from $1.66\times 10^5$ to $9.21\times 10^5~\mathrm{J\,m^{-2}\,s^{-1}\,T^{-1}}$ (a factor of $\sim$5.5) decreases the average AS from 24.37 to 14.78\,$R_\bigstar$ (39.4\%).

We also find that the highest $\frac{S}{B}$ scenario produces the highest radio emission frequencies for both HD~179949 and HD~189733~(\url{https://doi.org/10.25935/4pks-d207} \citep{Chebly2025_SI}, Figures C.2 and C.3). In particular, vertical emission~(those reaching high frequencies) become more prominent as $\frac{S}{B}$= increases. This likely results from an optimal balance between magnetic energy input and stellar wind conditions that favors emission generation. A higher~$\frac{S}{B}$ strengthens the stellar wind, causing the AS to move closer to the star where the magnetic field is stronger. Because the $f_{\rm ce }$ increases with magnetic field strength, a stronger $\frac{S}{B}$ naturally leads to higher-frequency radio emissions.

These results demonstrate that coronal properties, such as density and temperature, which directly affect the $\frac{S}{B}$ have a strong impact on the radio emission signature. Using a coronal model tailored to each star provides a more realistic representation of the system and helps avoid under- or overestimating the predicted radio emission frequencies.  

\titlespacing*{\section}{0pt}{6pt}{1pt} % 

\section{Star systems with sub-Alfv\'enic planet}
\label{app:Star systems from star sample}

We assess whether the planets in the six remaining star systems in \cite{Chebly2023}; HD~73256, HD~130322, $\epsilon$~Eri, HD~219134, DT~Vir, and HN~Peg lie within the AS under different $\frac{S}{B}$ scenarios, defining plausible $\frac{S}{B}$ ranges as in Section~\ref{fig:Sflux_eploration}.  
For most systems, $a \gg \overline{AS}_{\rm eq}$ (Table~\ref{tab:star_info}), so the AS remains well within the orbit even for $\frac{S}{B}$ as low as $10^{4}$--$10^{5}$. 
% For example, in HN~Peg decreasing $S/B$ by a factor of five increases $\overline{AS}_{\rm eq}$ by $\sim 30\%$ to $17.745~R_\bigstar$, still far smaller than $a = 1.60\times 10^5~R_\bigstar$.  
On the other hand for HD~130322 and HD~219134, $a$ is comparable to $\overline{AS}_{\rm eq}$. For HD~130322, $\overline{AS}_{\rm eq} = 13.40~R_\bigstar$ with $a = 24.0644~R_\bigstar$, and for HD~219134, $\overline{AS}_{\rm eq} = 8.33~R_\bigstar$ with $a_{\rm b} = 10.759~R_\bigstar$ and $a_{\rm c} = 17.822~R_\bigstar$. Their relatively low $L_{\rm X}$ imply $\frac{S}{B}$ ranges of $\sim 10^{2}$--$10^{3}$ and $10^{3}$--$10^{4}~\mathrm{J\,m^{-2}\,s^{-1}\,T^{-1}}$, resulting in a larger AS and making these systems candidates for possible detection of SPI-induced radio emission.

%\titlespacing*{\subsubsection}{0pt}{2pt}{1pt} % {left}{before}{after}
\section{Telescope sensitivity}
\label{app:Telescope sensitivity}

We adopt for all telescopes an integration time of 1 hour, a bandwidth of $\Delta \nu = 10$\,MHz, and two polarization channels ($n_{\rm p} = 2$). These values of the integration time and bandwidth allow for possible signal detection while remaining physically reasonable.

\subsubsection*{NenuFAR (10--85~MHz)}
NenuFAR operates from 10 to 85~MHz. At these frequencies, sky noise dominates over instrumental noise ($T_{\rm sys} \approx T_{\rm sky}$), with $T_{\rm sky} = 60 \, \lambda^{2.55}$\,K \citep{Thompson2017}, where $\lambda$ is in meters. Moreover, $A_{\rm eff} = 4452.10$\,m$^2$ at 10~MHz and 410.213\,m$^2$ at 85~MHz\,(\href{https://nenufar.obs-nancay.fr/en/astronomer/}{nenufar documentation}, \citealt{Zarka2020}).

\subsubsection*{LOFAR (30--80~MHz)}
LOFAR covers 10–240~MHz with multiple array configurations. For this study, we use the Full EU array, which provides an optimal combination of sensitivity and spatial resolution for observing star-planet systems \citep{vanHaarlem2013}. The SEFD for this array ranges from 6.5\,mJy at 30~MHz to 7.7\,mJy at 80~MHz, derived from Table~B.3 of \citet{vanHaarlem2013}. 

\subsubsection*{SKA1-Low (50--350~MHz)}
The Square Kilometre Array (SKA) comprises SKA1-Low (50–350,MHz, Western Australia) and SKA1-Mid (350,MHz–several GHz, South Africa) \citep{Dewdney2009}. We focus on SKA1-Low, optimal for detecting low-frequency radio emissions from star-planet interactions. We do not consider SKA2~(the planned upgrade), since its detailed technical specifications and performance parameters remain incomplete. Using parameters from the AA4 configuration\footnote{Aperture Array configuration number 4, which is one of the proposed layouts of the low-frequency antenna stations in SKA1-Low.} reported in the \href{https://www.skao.int/sites/default/files/documents/SKAO-TEL-0000818-V2_SKA1_Science_Performance.pdf}{SKA1 Science Performance document},~Table~5,~p.49, the effective area to system temperature ratio~($A_{\rm eff}/T_{\rm sys}$) ranges from 52.2\,m$^{2}$\,K$^{-1}$ at 50\,MHz to 591.8\,m$^{2}$\,K$^{-1}$ at 350\,MHz.

Table~\ref{tab:all_telescopes_parameters} summarizes the sensitivities and corresponding radio powers for each telescope across the target systems. As expected, the hierarchy in sensitivity follows SKA1-Low $>$ LOFAR $\gtrsim$ NenuFAR, with SKA1-Low being 1.5–2 orders of magnitude more sensitive. This implies that SKA1-Low can detect weaker radio emissions, while NenuFAR and LOFAR will require stronger signals to reach detection thresholds.
\begin{table*}[hbt!]
\centering
\caption{Parameter ranges for Tau~Boo, HD~179949, and HD~189733 observed with different telescopes.}
\resizebox{0.58\textwidth}{!}{%
\begin{tabular}{lcccc}
\toprule
\textbf{Telescope}  & \textbf{Parameter} & & \textbf{Values}& \\
\midrule
\multirow{4}{*}{\textbf{NenuFAR~(10-85~MHz)}} 
& SEFD\,(Jy)          & \multicolumn{3}{c}{$4.10\times 10^{2}$ - $4.452\times 10^{3} $ } \\
& $\sigma$\,(mJy)     & \multicolumn{3}{c}{$1.529\times 10^{-3}$ - $1.659\times 10^{-2}$} \\
& $S_{\min}$ (mJy)   & \multicolumn{3}{c}{$7.644\times 10^{-3}$ - $8.296\times 10^{-2}$} \\
% \multirow{4}{*}{} \\
&&\textbf{Tau~Boo} & \textbf{HD~179949} & \textbf{HD~189733}\\
& $P_{\mathrm{radio}}$\,(W) & $2.834\times 10^{13}$ – $3.076\times 10^{14}$ & $8.491\times 10^{13}$ - $9.215\times 10^{14}$& $4.339\times 10^{13}-4.709\times 10^{14}$\\

\midrule
\multirow{4}{*}{\textbf{LOFAR (30--80\,MHz)}} 
& SEFD~(Jy)          & \multicolumn{3}{c}{$1.74\times 10^{3}$ -- $2.06\times 10^{3}$} \\
& $\sigma$~(mJy)     & \multicolumn{3}{c}{6.5 -- 7.7} \\
& $S_{\min}$~(mJy)   & \multicolumn{3}{c}{32.5 -- 38.5} \\
% \multirow{4}{*}{} \\
&&\textbf{Tau~Boo} & \textbf{HD~179949} & \textbf{HD~189733}\\
& $P_{\mathrm{radio}}$\,(W) & $1.205\times 10^{14} - 1.428\times 10^{14}$ & $3.610\times 10^{14} - 4.277\times 10^{14}$ & $1.845\times 10^{14}- 2.185\times 10^{14}$  \\
\midrule
\multirow{4}{*}{\textbf{SKA1-Low (50--350 MHz)}} 
& SEFD\,(Jy)          & \multicolumn{3}{c}{4.67 -- 52.90} \\
& $\sigma$\,(mJy)     & \multicolumn{3}{c}{$1.739 \times 10^{-2} - 0.19$} \\
& $S_{\min}$\,(mJy)   & \multicolumn{3}{c}{$8.69\times 10^{-2} - 9.86\times 10^{-1}$} \\
% \multirow{4}{*}{} \\
&&\textbf{Tau~Boo} & \textbf{HD~179949} & \textbf{HD~189733}\\
& $P_{\mathrm{radio}}$\,(W) & $3.224\times 10^{11} - 3.655\times 10^{12}$ & $9.658\times 10^{11} - 1.095\times 10^{13}$ & $4.935\times 10^{11} - 5.595\times 10^{12}$\\

\bottomrule
\end{tabular}%
}
\vspace{0.3em}
\tablefoot{For each telescope, we provide the system equivalent flux density (SEFD), the detection threshold ($\sigma$), the minimum detectable flux density ($S_{\rm min}$), and the corresponding range of minimum detectable radio power ($P_{\rm radio}$) for Tau~Boo, HD~179949, and HD~189733.}
\label{tab:all_telescopes_parameters}
\end{table*}

\section{Maximum elevation of target stars}
\label{app:max_elevation}

The elevation of a star above the local horizon determines its observability with a given radio telescope. We focus here on the maximum elevation, $\alpha_{\rm max}$, which is the angular height a star attains when crossing the local meridian~(its highest point in the sky). This quantity is independent of the observation date.
The maximum elevation depends only on the observer’s latitude ($\phi_0$) and the star’s declination ($\delta$) and can be computed via the standard equatorial-to-horizontal coordinate transformation:
\begin{equation}
\alpha = \arcsin \Big( \sin\phi_0 \times \sin\delta + \cos\phi_0 \times \cos\delta \times \cos h \Big),
\end{equation}
where $h$ is the hour angle of the star (Local Sidereal Time - Right Ascension). The maximum elevation occurs at $h = 0$, i.e., when the star crosses the meridian.

The declinations of the target stars were obtained from \href{https://simbad.u-strasbg.fr/simbad/}{SIMBAD}, while telescope latitudes correspond to the observatories' geographic locations. Applying the above formula, we computed the maximum elevations for each star-telescope pair. 
The target stars considered are Tau Boo~($\delta = 17.5^\circ$), HD~179949 ($\delta = -24.8^\circ$), and HD 189733 ($\delta = 22.7^\circ$). The telescope latitudes are:~NenuFAR~(47.38$^\circ$, Nançay,~France), LOFAR~(52.91$^\circ$,~Netherlands), and SKA1-Low~(-26.7$^\circ$,~Western Australia). These results are summarized in Table~\ref{tab:star_tel_coords_elev}. 

\begin{table}[h!]
\centering
\caption{Maximum elevation $\alpha_{\rm max}$ of each target.}
\label{tab:star_tel_coords_elev}
\resizebox{0.6\columnwidth}{!}{
\begin{tabular}{lcccc}
\hline
Star & NenuFAR ($^\circ$) & LOFAR ($^\circ$) & SKA1-Low ($^\circ$)  \\
\hline
Tau Boo       & 60.12 & 54.59 & 45.80  \\
HD 179949     & 17.92 & 12.39 & 88.00  \\
HD 189733     & 65.32 & 59.79 & 40.60 \\
\hline
\end{tabular}
}
\end{table}


\begin{thebibliography}{}

\bibitem[\protect\citeauthoryear{Acharya et al.}{2023}]{Acharya2023} Acharya A., Kashyap V.~L., Saar S.~H., Singh K.~P., Cuntz M., 2023, ApJ, 951, 152. doi:10.3847/1538-4357/acd054

\bibitem[\protect\citeauthoryear{Alvarado-G{\'o}mez et al.}{2015}]{Alvarado2015} Alvarado-G{\'o}mez J.~D., Hussain G.~A.~J., Grunhut J., Fares R., Donati J.-F., Alecian E., Kochukhov O., et al., 2015, A\&A, 582, A38. doi:10.1051/0004-6361/201525771

\bibitem[\protect\citeauthoryear{Alvarado-G{\'o}mez et al.}{2016}]{Alvarado2016} Alvarado-G{\'o}mez J.~D., Hussain G.~A.~J., Cohen O., Drake J.~J., Garraffo C., Grunhut J., Gombosi T.~I., 2016, A\&A, 594, A95. doi:10.1051/0004-6361/201628988

\bibitem[\protect\citeauthoryear{Alvarado-G{\'o}mez et al.}{2018}]{Alvarado2018} Alvarado-G{\'o}mez J.~D., Drake J.~J., Cohen O., Moschou S.~P., Garraffo C., 2018, ApJ, 862, 93. doi:10.3847/1538-4357/aacb7f

\bibitem[\protect\citeauthoryear{Alvarado-G{\'o}mez et al.}{2019}]{Alvarado2019} Alvarado-G{\'o}mez J.~D., Drake J.~J., Moschou S.~P., Garraffo C., Cohen O., NASA LWS Focus Science Team: Solar-Stellar Connection, Yadav R.~K., et al., 2019, ApJL, 884, L13. doi:10.3847/2041-8213/ab44d0

\bibitem[\protect\citeauthoryear{Alvarado-G\'omez et al.}{2020}]{AlvaradoGomes2020} 
Alvarado-G\'omez J.~D., Drake J.~J., Garraffo C., Cohen O., Moschou S.-P., Fraschetti F., 2020, ApJ, 895, 47. \url{https://doi.org/10.3847/1538-4357/ab8941}

% \bibitem[\protect\citeauthoryear{Ashtari et al.}{2022}]{Ashtari2022} Ashtari R., Sciola A., Turner J.~D., Stevenson K., 2022, ApJ, 939, 24. doi:10.3847/1538-4357/ac92f5


% \bibitem[\protect\citeauthoryear{Bastian, Dulk, \& Leblanc}{2000}]{Bastian2000} Bastian T.~S., Dulk G.~A., Leblanc Y., 2000, ApJ, 545, 1058. doi:10.1086/317864

\bibitem[\protect\citeauthoryear{Bigg}{1964}]{Bigg1964} Bigg E.~K., 1964, Natur, 203, 1008. doi:10.1038/2031008a0

% \bibitem[\protect\citeauthoryear{Bloot et al.}{2024}]{Bloot2024} Bloot S., Callingham J.~R., Vedantham H.~K., Kavanagh R.~D., Pope B.~J.~S., Climent J.~B., Guirado J.~C., et al., 2024, A\&A, 682, A170. doi:10.1051/0004-6361/202348065

% \bibitem[\protect\citeauthoryear{Borsa et al.}{2014}]{Borsa2014} Borsa, F., Poretti, E., Rainer, M., et al., 2014, Star–planet interaction in the HD~179949 system, \emph{Astronomy \& Astrophysics}, 565, A112. \url{https://www.aanda.org/articles/aa/pdf/2014/05/aa23537-14.pdf}

\bibitem[\protect\citeauthoryear{Brogi et al.}{2012}]{Brogi2012} Brogi M., Snellen I.~A.~G., de Kok R.~J., Albrecht S., Birkby J., de Mooij E.~J.~W., 2012, Natur, 486, 502. doi:10.1038/nature11161

\bibitem[\protect\citeauthoryear{Brogi et al.}{2014}]{Brogi2014} Brogi M., de Kok R.~J., Birkby J.~L., Schwarz H., Snellen I.~A.~G., 2014, A\&A, 565, A124. doi:10.1051/0004-6361/201423537

\bibitem[\protect\citeauthoryear{Butler et al.}{2006}]{Butler2006} Butler, R. P., Wright, J. T., Marcy, G. W., et al., 2006, Catalog of Nearby Exoplanets, \emph{The Astrophysical Journal}, 646(1), 505–522.

\bibitem[\protect\citeauthoryear{Callingham et al.}{2021}]{Callingham2021} Callingham J.~R., Vedantham H.~K., Shimwell T.~W., Pope B.~J.~S., Davis I.~E., Best P.~N., Hardcastle M.~J., et al., 2021, NatAs, 5, 1233. doi:10.1038/s41550-021-01483-0

\bibitem[\protect\citeauthoryear{Callingham et al.}{2024}]{Callingham2024} Callingham J.~R., Pope B.~J.~S., Kavanagh R.~D., Bellotti S., Daley-Yates S., Damasso M., Grie{\ss}meier J.-M., et al., 2024, NatAs, 8, 1359. doi:10.1038/s41550-024-02405-6

\bibitem[\protect\citeauthoryear{Cecconi et al.}{2012}]{Cecconi2012} Cecconi B., Hess S., H\'erique A., Santovito M.~R., Santos-Costa D., Zarka P., Alberti G.,  
Blankenship D., Bougeret J.~L., Bruzzone L., Kofman w., 2012, PSS, 61, 32. 
doi = 10.1016/j.pss.2011.06.012

\bibitem[\protect\citeauthoryear{Cauley et al.}{2019}]{Cauley2019} Cauley P.~W., Shkolnik E.~L., Llama J., Lanza A.~F., 2019, NatAs, 3, 1128. doi:10.1038/s41550-019-0840-x

\bibitem[\protect\citeauthoryear{Chebly, Alvarado-G{\'o}mez, \& Poppenhaeger}{2022}]{Chebly2022} Chebly J.~J., Alvarado-G{\'o}mez J.~D., Poppenhaeger K., 2022, AN, 343, e10093. doi:10.1002/asna.20210093

\bibitem[\protect\citeauthoryear{Chebly et al.}{2023}]{Chebly2023} Chebly J.~J., Alvarado-G{\'o}mez J.~D., Poppenh{\"a}ger K., Garraffo C., 2023, MNRAS, 524, 5060. doi:10.1093/mnras/stad2100

\bibitem[\protect\citeauthoryear{Chebly et al.}{2025a}]{Chebly2025_SI}Chebly, J. J., Louis, C. K., Strugarek, A., Alvarado-G{\'o}mez J.~D., Zarka  P., 2025a, PADC, doi: 10.25935/4pks-d207

\bibitem[\protect\citeauthoryear{Chebly et al.}{2025b}]{Chebly2025_ExPRES_TauBoo}
Chebly, J. J., Louis, C. K., Strugarek, A, 2025b, PADC, doi: 10.25935/9c80-w202

\bibitem[\protect\citeauthoryear{Chebly et al.}{2025c}]{Chebly2025_ExPRES_HD189733}
Chebly, J. J., Louis, C. K., Strugarek, A., 2025c, PADC, doi: 10.25935/ce9q-n094

\bibitem[\protect\citeauthoryear{Chebly et al.}{2025d}]{Chebly2025_ExPRES_HD179949}
Chebly, J. J., Louis, C. K., Strugarek, A., 2025d, PADC, doi: 10.25935/d1h2-z977

% \bibitem[\protect\citeauthoryear{Cohen et al.}{2007}]{Cohen2007} Cohen O., Sokolov I.~V., Roussev I.~I., Arge C.~N., Manchester W.~B., Gombosi T.~I., Frazin R.~A., et al., 2007, ApJL, 654, L163. doi:10.1086/511154

% \bibitem[\protect\citeauthoryear{Cohen \& Drake}{2014}]{Cohen2014} Cohen O., Drake J.~J., 2014, ApJ, 783, 55. doi:10.1088/0004-637X/783/1/55

\bibitem[\protect\citeauthoryear{Cohen et al.}{2020}]{Cohen2020} Cohen O., Garraffo C., Moschou S.-P., Drake J.~J., Alvarado-G{\'o}mez J.~D., Glocer A., Fraschetti F., 2020, ApJ, 897, 101. doi:10.3847/1538-4357/ab9637

\bibitem[\protect\citeauthoryear{Colombo et al.}{2024}]{Colombo2024} Colombo S., Pillitteri I., Petralia A., Orlando S., Micela G., 2024, A\&A, 683, A226. doi:10.1051/0004-6361/202348289


\bibitem[\protect\citeauthoryear{Cristofari et al.}{2022}]{Cristofari2022} Cristofari P.~I., Donati J.-F., Masseron T., Fouqu{\'e} P., Moutou C., Carmona A., Artigau E., et al., 2022, MNRAS, 516, 3802. doi:10.1093/mnras/stac2364

\bibitem[\protect\citeauthoryear{Desch}{1982}]{Desch1982} Desch M.~D., 1982, JGR, 87, 4549. doi:10.1029/JA087iA06p04549

\bibitem[\protect\citeauthoryear{Desch}{1983}]{Desch1983} Desch M.~D., 1983, JGR, 88, A9. doi:10.1029/JA088iA09p06904


\bibitem[\protect\citeauthoryear{Desch \& Kaiser}{1984}]{Desch1984} Desch M.~D., Kaiser M.~L., 1984, Natur, 310, 755. doi:10.1038/310755a0

\bibitem[\protect\citeauthoryear{Dewdney et al.}{2009}]{Dewdney2009} 
Dewdney P.~E., Hall P.~J., Schilizzi R.~T., Lazio T.~J.~W., 2009, Proc. IEEE, 97, 1482. doi.org/10.1109/JPROC.2009.2021005

\bibitem[\protect\citeauthoryear{Donati et al.}{2006}]{Donati2006} Donati J.-F., Howarth I.~D., Jardine M.~M., Petit P., Catala C., Landstreet J.~D., Bouret J.-C., et al., 2006, MNRAS, 370, 629. doi:10.1111/j.1365-2966.2006.10558.x

\bibitem[\protect\citeauthoryear{Dulk}{1985}]{Dulk1985} Dulk G.~A., 1985, ARA\&A, 23, 169. doi:10.1146/annurev.aa.23.090185.001125

\bibitem[\protect\citeauthoryear{Fares et al.}{2009}]{Fares2009} Fares R., Donati J.-F., Moutou C., Bohlender D., Catala C., Deleuil M., Shkolnik E., et al., 2009, MNRAS, 398, 1383. doi:10.1111/j.1365-2966.2009.15303.x

\bibitem[\protect\citeauthoryear{Fares et al.}{2010}]{Fares2010} 
Fares R., Donati J.-F., Moutou C., Jardine M.~M., Grießmeier J.-M., Zarka P., Shkolnik E., et al., 2010, MNRAS, 406, 409. doi.org/10.1111/j.1365-2966.2010.16691.x

\bibitem[\protect\citeauthoryear{Fares et al.}{2012}]{Fares2012} Fares R., Donati J.-F., Moutou C., Jardine M., Cameron A.~C., Lanza A.~F., Bohlender D., et al., 2012, MNRAS, 423, 1006. doi:10.1111/j.1365-2966.2012.20780.x

\bibitem[\protect\citeauthoryear{Fares et al.}{2013}]{Fares2013} Fares R., Moutou C., Donati J.-F., Catala C., Shkolnik E.~L., Jardine M.~M., Cameron A.~C., et al., 2013, MNRAS, 435, 1451. doi:10.1093/mnras/stt1386

\bibitem[\protect\citeauthoryear{Farrell, Desch, \& Zarka}{1999}]{Farrell1999} Farrell W.~M., Desch M.~D., Zarka P., 1999, JGR, 104, 14025. doi:10.1029/1998JE900050

\bibitem[\protect\citeauthoryear{Finley et al.}{2020}]{Finley2020} Finley A.~J., Matt S.~P., R{\'e}ville V., Pinto R.~F., Owens M., Kasper J.~C., Korreck K.~E., et al., 2020, ApJL, 902, L4. doi:10.3847/2041-8213/abb9a5

\bibitem[\protect\citeauthoryear{Fischer \& Saur}{2019}]{Fischer2019} Fischer C., Saur J., 2019, ApJ, 872, 113. doi:10.3847/1538-4357/aafaf2

\bibitem[\protect\citeauthoryear{Gaia Collaboration et al.}{2021}]{Gaia2021} Gaia Collaboration, Brown, A. G. A., Vallenari, A., et al., 2021, Gaia Early Data Release 3, \emph{Astronomy \& Astrophysics}, 649, A1.

\bibitem[\protect\citeauthoryear{Gallagher \& Dangelo}{1981}]{Gallagher1981} Gallagher D.~L., Dangelo N., 1981, GeoRL, 8, 1087. doi:10.1029/GL008i010p01087

\bibitem[\protect\citeauthoryear{Garraffo et al.}{2017}]{Garraffo2017} Garraffo C., Drake J.~J., Cohen O., Alvarado-G{\'o}mez J.~D., Moschou S.~P., 2017, ApJL, 843, L33. doi:10.3847/2041-8213/aa79ed

\bibitem[\protect\citeauthoryear{Gombosi et al.}{2004}]{Gombosi2004} Gombosi T.~I., Powell K.~G., De Zeeuw D.~L., Clauer C.~R., Hansen K.~C., Manchester W.~B., Ridley A.~J., et al., 2004, CSE, 6, 14. doi:10.1109/MCISE.2004.1267603

\bibitem[\protect\citeauthoryear{Gombosi et al.}{2018}]{Gombosi2018} Gombosi T.~I., van der Holst B., Manchester W.~B., Sokolov I.~V., 2018, LRSP, 15, 4. doi:10.1007/s41116-018-0014-4

\bibitem[\protect\citeauthoryear{Gupta et al.}{2017}]{Gupta2017} Gupta Y., Ajithkumar B., Kale H.~S., Nayak S., Sabhapathy S., Sureshkumar S., Swami R.~V., et al., 2017, CSci, 113, 707. doi:10.18520/cs/v113/i04/707-714

\bibitem[\protect\citeauthoryear{Hahlin et al.}{2023}]{Hahlin2023} Hahlin A., Kochukhov O., Rains A.~D., Lavail A., Hatzes A., Piskunov N., Reiners A., et al., 2023, A\&A, 675, A91. doi:10.1051/0004-6361/202346314

\bibitem[\protect\citeauthoryear{Hazra et al.}{2021}]{Hazra2021} Hazra S., R{\'e}ville V., Perri B., Strugarek A., Brun A.~S., Buchlin E., 2021, ApJ, 910, 90. doi:10.3847/1538-4357/abe12e


\bibitem[\protect\citeauthoryear{Hess et al.}{2007}]{hess2007} 
Hess, S. L. G., Mottez, F., Zarka, P., 2007, Generation of the Jovian decameter radio emissions: Statistical study of Io-related emissions using the shell-type electron distribution, Journal of Geophysical Research: Space Physics, 112, A11. doi.org/10.1029/2007JA012308

\bibitem[\protect\citeauthoryear{Hess, Cecconi, \& Zarka}{2008}]{Hess2008a} Hess S., Cecconi B., Zarka P., 2008, GeoRL, 35, L13107. doi:10.1029/2008GL033656

\bibitem[\protect\citeauthoryear{Hess et al.}{2011}]{Hess2011} Hess S.~L.~G., Bonfond B., Zarka P., Grodent D., 2011, JGRA, 116, A05217. doi:10.1029/2010JA016262

\bibitem[\protect\citeauthoryear{Hilgers et al.}{1992}]{Hilgers1992} Hilgers A., Holback B., Holmgren G., Bostrom R., 1992, JGR, 97, 8631. doi:10.1029/91JA02193


\bibitem[\protect\citeauthoryear{Hori}{2021}]{Hori2021} Hori Y., 2021, ApJ, 908, 77. doi:10.3847/1538-4357/abd8d1

\bibitem[\protect\citeauthoryear{Hue et al.}{2023}]{Hue2023} Hue V., Gladstone G.~R., Louis C.~K., Greathouse T.~K., Bonfond B., Szalay J.~R., Moirano A., et al., 2023, JGRA, 128, e2023JA031363. doi:10.1029/2023JA031363

\bibitem[\protect\citeauthoryear{Hussain et al.}{2016}]{Hussain2016} Hussain G.~A.~J., Alvarado-G{\'o}mez J.~D., Grunhut J., Donati J.-F., Alecian E., Oksala M., Morin J., et al., 2016, A\&A, 585, A77. doi:10.1051/0004-6361/201526595

\bibitem[\protect\citeauthoryear{Ilin et al.}{2025}]{Ilin2025} Ilin E., Bloot S., Callingham J.~R., Vedantham H.~K., 2025, A\&A, 699, A147. doi:10.1051/0004-6361/202554684


\bibitem[\protect\citeauthoryear{J{\'a}come et al.}{2022}]{Jacome2022} J{\'a}come H.~R.~P., Marques M.~S., Zarka P., Echer E., Lamy L., Louis C.~K., 2022, A\&A, 665, A67. doi:10.1051/0004-6361/202244246

\bibitem[\protect\citeauthoryear{Johnston et al.}{2008}]{Johnston2008} Johnston S., Taylor R., Bailes M., Bartel N., Baugh C., Bietenholz M., Blake C., et al., 2008, ExA, 22, 151. doi:10.1007/s10686-008-9124-7

\bibitem[\protect\citeauthoryear{Justesen \& Albrecht}{2019}]{Justesen2019} Justesen A.~B., Albrecht S., 2019, A\&A, 625, A59. doi:10.1051/0004-6361/201834368


\bibitem[\protect\citeauthoryear{Kaiser et al.}{2000}]{Kaiser2000} Kaiser M.~L., Zarka P., Kurth W.~S., Hospodarsky G.~B., Gurnett D.~A., 2000, JGR, 105, 16053. doi:10.1029/1999JA000414

\bibitem[\protect\citeauthoryear{Kavanagh et al.}{2019}]{Kavanagh2019} Kavanagh R.~D., Vidotto A.~A., {\'O}. Fionnag{\'a}in D., Bourrier V., Fares R., Jardine M., Helling C., et al., 2019, MNRAS, 485, 4529. doi:10.1093/mnras/stz655

\bibitem[\protect\citeauthoryear{Kavanagh et al.}{2021}]{Kavanagh2021} Kavanagh R.~D., Vidotto A.~A., Klein B., Jardine M.~M., Donati J.-F., {\'O} Fionnag{\'a}in D., 2021, csss.conf, 315. doi:10.5281/zenodo.4728000

\bibitem[\protect\citeauthoryear{Kavanagh \& Vedantham}{2023}]{Kavanagh2023} Kavanagh R.~D., Vedantham H.~K., 2023, MNRAS, 524, 6267. doi:10.1093/mnras/stad2035

\bibitem[\protect\citeauthoryear{Kochukhov \& Piskunov}{2002}]{Kochukhov2002} Kochukhov O., Piskunov N., 2002, A\&A, 388, 868. doi:10.1051/0004-6361:20020300

\bibitem[\protect\citeauthoryear{Kochukhov et al.}{2020}]{Kochukhov2020} Kochukhov O., Hackman T., Lehtinen J.~J., Wehrhahn A., 2020, A\&A, 635, A142. doi:10.1051/0004-6361/201937185

\bibitem[\protect\citeauthoryear{Kokori et al.}{2023}]{Kokori2023} Kokori A., Tsiaras A., Edwards B., Jones A., Pantelidou G., Tinetti G., Bewersdorff L., et al., 2023, ApJS, 265, 4. doi:10.3847/1538-4365/ac9da4

\bibitem[\protect\citeauthoryear{Lang et al.}{2014}]{Lang2014} Lang P., Jardine M., Morin J., Donati J.-F., Jeffers S., Vidotto A.~A., Fares R., 2014, MNRAS, 439, 2122. doi:10.1093/mnras/stu091

\bibitem[\protect\citeauthoryear{Lamy et al.}{2008}]{Lamy2008a} Lamy L., Zarka P., Cecconi B., Hess S., Prang{\'e} R., 2008, JGRA, 113, A10213. doi:10.1029/2008JA013464

\bibitem[\protect\citeauthoryear{Lamy et al.}{2013}]{Lamy2013} Lamy L., Prang{\'e} R., Pryor W., Gustin J., Badman S.~V., Melin H., Stallard T., et al., 2013, JGRA, 118, 4817. doi:10.1002/jgra.50404

\bibitem[\protect\citeauthoryear{Lamy et al.}{2022}]{Lamy2022} Lamy L., Colomban L., Zarka P., Prang{\'e} R., Marques M.~S., Louis C.~K., Kurth W.~S., et al., 2022, JGRA, 127, e30160. doi:10.1029/2021JA030160

\bibitem[\protect\citeauthoryear{Lamy et al.}{2023}]{Lamy2023} Lamy L., Waters J.~E., Louis C.~K., 2023, pre9.conf, 103091. doi:10.25546/103091

\bibitem[\protect\citeauthoryear{Lanza}{2012}]{Lanza2012} Lanza A.~F., 2012, A\&A, 544, A23. doi:10.1051/0004-6361/201219002

\bibitem[\protect\citeauthoryear{Lazio et al.}{2004}]{Lazio2004} Lazio T.~J., Farrell W.~M., Dietrick J., Greenlees E., Hogan E., Jones C., Hennig L.~A., 2004, ApJ, 612, 511. doi:10.1086/422449

\bibitem[\protect\citeauthoryear{Lazio et al.}{2010}]{Lazio2010} Lazio T.~J.~W., Carmichael S., Clark J., Elkins E., Gudmundsen P., Mott Z., Szwajkowski M., et al., 2010, AJ, 139, 96. doi:10.1088/0004-6256/139/1/96

\bibitem[\protect\citeauthoryear{Lazio}{2024}]{Lazio2024} Lazio T.~J.~W., 2024, arXiv, arXiv:2404.12348. doi:10.48550/arXiv.2404.12348

\bibitem[\protect\citeauthoryear{Louis et al.}{2017}]{Louis2017a} Louis C.~K., Lamy L., Zarka P., Cecconi B., Imai M., Kurth W.~S., Hospodarsky G., et al., 2017, GeoRL, 44, 9225. doi:10.1002/2017GL073036

\bibitem[\protect\citeauthoryear{Louis et al.}{2017}]{Louis2017} Louis C.~K., Lamy L., Zarka P., Cecconi B., Hess S.~L.~G., 2017, JGRA, 122, 9228. doi:10.1002/2016JA023779

\bibitem[\protect\citeauthoryear{Louis et al.}{2019}]{Louis2019a} Louis C.~K., Hess S.~L.~G., Cecconi B., Zarka P., Lamy L., Aicardi S., Loh A., 2019, A\&A, 627, A30. doi:10.1051/0004-6361/201935161

\bibitem[\protect\citeauthoryear{Louis et al.}{2019}]{Louis2019} 
Louis C.~K., Cecconi B., Hess S.~L.~G., Zarka P., 2019, A\&A, 627, A30. \url{https://doi.org/10.1051/0004-6361/201935329}

\bibitem[\protect\citeauthoryear{Louis et al.}{2020}]{Louis2020} Louis C., Louarn P., Allegrini F., Kurth W.~S., Szalay J.~R., 2020, AGUFMSM04, 2020, SM049-08

\bibitem[\protect\citeauthoryear{Louis et al.}{2023a}]{Louis2023} Louis C.~K., Louarn P., Collet B., Cl{\'e}ment N., Al Saati S., Szalay J.~R., Hue V., et al., 2023, JGRA, 128, e2023JA031985. doi:10.1029/2023JA031985

\bibitem[\protect\citeauthoryear{Louis et al.}{2023b}]{Louis2023b}
Louis, C. K., Jackman, C. M., Hospodarsky, G., O’Kane Hackett, A., Devon-Hurley, E., Zarka, P., et al., 2023, Journal of
Geophysical Research: Space Physics,
128, e2022JA031155. doi: 10.1029/2022JA031155

\bibitem[\protect\citeauthoryear{Louis et al.}{2025}]{Louis2025}
Louis, C. K., Hess, S. L. G., Cecconi, B., Zarka, P., Lamy, L., Aicardi, S., Loh, A., 2025, Zenodo, doi: 10.5281/zenodo.17047296

\bibitem[\protect\citeauthoryear{Mauduit et al.}{2023}]{Mauduit2023} Mauduit E., Zarka P., Lamy L., Hess S.~L.~G., 2023, NatCo, 14, 5981. doi:10.1038/s41467-023-41617-8

\bibitem[\protect\citeauthoryear{Mignone et al.}{2007}]{Mignone2007} Mignone A., Bodo G., Massaglia S., Matsakos T., Tesileanu O., Zanni C., Ferrari A., 2007, ApJS, 170, 228. doi:10.1086/513316

\bibitem[\protect\citeauthoryear{Morin et al.}{2008}]{Morin2008} Morin J., Donati J.-F., Petit P., Delfosse X., Forveille T., Albert L., Auri{\`e}re M., et al., 2008, MNRAS, 390, 567. doi:10.1111/j.1365-2966.2008.13809.x

\bibitem[\protect\citeauthoryear{Nan}{2006}]{Nan2006} Nan R., 2006, ScChG, 49, 129. doi:10.1007/s11433-006-0129-9

\bibitem[\protect\citeauthoryear{Nan et al.}{2011}]{Nan2011} Nan R., Li D., Jin C., Wang Q., Zhu L., Zhu W., Zhang H., et al., 2011, IJMPD, 20, 989. doi:10.1142/S0218271811019335

\bibitem[\protect\citeauthoryear{Neubauer}{1980}]{Neubauer1980} Neubauer F.~M., 1980, JGR, 85, 1171. doi:10.1029/JA085iA03p01171

\bibitem[\protect\citeauthoryear{Nichols \& Milan}{2016}]{Nichols2016} Nichols J.~D., Milan S.~E., 2016, MNRAS, 461, 2353. doi:10.1093/mnras/stw1430

\bibitem[\protect\citeauthoryear{Pe{\~n}a-Mo{\~n}ino et al.}{2024}]{2024A&A...688A.138P} Pe{\~n}a-Mo{\~n}ino L., P{\'e}rez-Torres M., Varela J., Zarka P., 2024, A\&A, 688, A138. doi:10.1051/0004-6361/202349042

\bibitem[\protect\citeauthoryear{Perri et al.}{2024}]{Perri2024} Perri B., Finley A., R{\'e}ville V., Parenti S., Brun A.~S., Strugarek A., Buchlin {\'E}., 2024, A\&A, 687, A10. doi:10.1051/0004-6361/202349040

\bibitem[\protect\citeauthoryear{Parenti et al.}{2022}]{Parenti2022} Parenti S., R{\'e}ville V., Brun A.~S., Pinto R.~F., Auch{\`e}re F., Buchlin {\'E}., Perri B., et al., 2022, ApJ, 929, 75. doi:10.3847/1538-4357/ac56da

\bibitem[\protect\citeauthoryear{Paul et al.}{2025}]{Paul2025} Paul A., Strugarek A., et al., 2025, A\&A, under review

\bibitem[\protect\citeauthoryear{Pe{\~n}a-Mo{\~n}ino et al.}{2024}]{Pena2024} Pe{\~n}a-Mo{\~n}ino L., P{\'e}rez-Torres M., Varela J., Zarka P., 2024, A\&A, 688, A138. doi:10.1051/0004-6361/202349042

\bibitem[\protect\citeauthoryear{P{\'e}rez-Torres et al.}{2021}]{Perez2021} P{\'e}rez-Torres M., G{\'o}mez J.~F., Ortiz J.~L., Leto P., Anglada G., G{\'o}mez J.~L., Rodr{\'\i}guez E., et al., 2021, A\&A, 645, A77. doi:10.1051/0004-6361/202039052

\bibitem[\protect\citeauthoryear{Perley et al.}{2011}]{Perley2011} Perley R.~A., Chandler C.~J., Butler B.~J., Wrobel J.~M., 2011, ApJL, 739, L1. doi:10.1088/2041-8205/739/1/L1

\bibitem[\protect\citeauthoryear{Piddington}{1977}]{Piddington1977} Piddington J.~H., 1977, Moon, 17, 373. doi:10.1007/BF00562646

\bibitem[\protect\citeauthoryear{Pineda \& Villadsen}{2023}]{Pineda2023} Pineda J.~S., Villadsen J., 2023, NatAs, 7, 569. doi:10.1038/s41550-023-01914-0

\bibitem[\protect\citeauthoryear{Pizzolato et al.}{2003}]{Pizzolato2003} Pizzolato N., Maggio A., Micela G., Sciortino S., Ventura P., 2003, A\&A, 397, 147. doi:10.1051/0004-6361:20021560

\bibitem[\protect\citeauthoryear{Powell et al.}{1999}]{Powell1999} Powell K.~G., Roe P.~L., Linde T.~J., Gombosi T.~I., De Zeeuw D.~L., 1999, JCoPh, 154, 284. doi:10.1006/jcph.1999.6299

\bibitem[\protect\citeauthoryear{Pritchard et al.}{2021}]{Pritchard2021} Pritchard J., Murphy T., Zic A., Lynch C., Heald G., Kaplan D.~L., Anderson C., et al., 2021, MNRAS, 502, 5438. doi:10.1093/mnras/stab299

\bibitem[\protect\citeauthoryear{Pritchett}{1986}]{Pritchett1986} Pritchett P.~L., 1986, PhFl, 29, 2919. doi:10.1063/1.865492

\bibitem[\protect\citeauthoryear{Queinnec \& Zarka}{1998}]{Queinnec1998} Queinnec J., Zarka P., 1998, JGR, 103, 26649. doi:10.1029/98JA02435

\bibitem[\protect\citeauthoryear{Ray \& Hess}{2008}]{Ray2008} Ray L.~C., Hess S., 2008, JGRA, 113, A11218. doi:10.1029/2008JA013669

\bibitem[\protect\citeauthoryear{Reville et al.}{2020}]{Reville2020} R{\'e}ville V., Velli M., Rouillard A.~P., Lavraud B., Tenerani A., Shi C., Strugarek A., 2020, ApJL, 895, L20. doi:10.3847/2041-8213/ab911d

\bibitem[\protect\citeauthoryear{R{\'e}ville et al.}{2024}]{Reville2024} R{\'e}ville V., Jasinski J.~M., Velli M., Strugarek A., Brun A.~S., Murphy N., Regoli L.~H., et al., 2024, ApJ, 976, 65. doi:10.3847/1538-4357/ad8132

\bibitem[\protect\citeauthoryear{Rosenthal et al.}{2021}]{Rosenthal2021} Rosenthal L.~J., Fulton B.~J., Hirsch L.~A., Isaacson H.~T., Howard A.~W., Dedrick C.~M., Sherstyuk I.~A., et al., 2021, ApJS, 255, 8. doi:10.3847/1538-4365/abe23c

\bibitem[\protect\citeauthoryear{Saur et al.}{2013}]{Saur2013} Saur J., Grambusch T., Duling S., Neubauer F.~M., Simon S., 2013, A\&A, 552, A119. doi:10.1051/0004-6361/201118179

\bibitem[\protect\citeauthoryear{Schmitt \& Liefke}{2004}]{Schmitt2004} Schmitt J.~H.~M.~M., Liefke C., 2004, A\&A, 417, 651. doi:10.1051/0004-6361:20030495

\bibitem[\protect\citeauthoryear{Shields, Ballard, \& Johnson}{2016}]{Shields2016} Shields A.~L., Ballard S., Johnson J.~A., 2016, PhR, 663, 1. doi:10.1016/j.physrep.2016.10.003

\bibitem[\protect\citeauthoryear{Sokolov et al.}{2013}]{Sokolov2013} Sokolov I.~V., van der Holst B., Oran R., Downs C., Roussev I.~I., Jin M., Manchester W.~B., et al., 2013, ApJ, 764, 23. doi:10.1088/0004-637X/764/1/23

\bibitem[\protect\citeauthoryear{Stassun, Collins, \& Gaudi}{2017}]{Stassun2017} Stassun K.~G., Collins K.~A., Gaudi B.~S., 2017, AJ, 153, 136. doi:10.3847/1538-3881/aa5df3

\bibitem[\protect\citeauthoryear{Strickert, Evensberget, \& Vidotto}{2024}]{Strickert2024} Strickert K.~M., Evensberget D., Vidotto A.~A., 2024, MNRAS, 533, 1156. doi:10.1093/mnras/stae1884

\bibitem[\protect\citeauthoryear{Strugarek et al.}{2015}]{Strugarek2015} Strugarek A., Brun A.~S., Matt S.~P., R{\'e}ville V., 2015, ApJ, 815, 111. doi:10.1088/0004-637X/815/2/111

\bibitem[\protect\citeauthoryear{Strugarek}{2016}]{Strugarek2016} Strugarek A., 2016, ApJ, 833, 140. doi:10.3847/1538-4357/833/2/140

\bibitem[\protect\citeauthoryear{Strugarek et al.}{2022}]{Strugarek2022} Strugarek A., Fares R., Bourrier V., Brun A.~S., R{\'e}ville V., Amari T., Helling C., et al., 2022, MNRAS, 512, 4556. doi:10.1093/mnras/stac778

\bibitem[\protect\citeauthoryear{Su}{2009}]{Su2009} Su Y.-J., 2009, IAUS, 259, 271. doi:10.1017/S1743921309030610

\bibitem[\protect\citeauthoryear{Swarup et al.}{1991}]{Swarup1991} Swarup G., Ananthakrishnan S., Kapahi V.~K., Rao A.~P., Subrahmanya C.~R., Kulkarni V.~K., 1991, CSci, 60, 95

\bibitem[\protect\citeauthoryear{Tasse, C., P. Zarka, M. Hardcastle, et al.}{2025}]{Tasse2025}, Tasse, C., P. Zarka, M. Hardcastle, et al., 2025, The detection of circularly polarized radio bursts from stellar and exoplanetary systems, Nature Astronomy, in revision, 2025

\bibitem[\protect\citeauthoryear{Thompson, Moran, \& Swenson}{2017}]{Thompson2017} Thompson A.~R., Moran J.~M., Swenson G.~W., 2017, isra.book. doi:10.1007/978-3-319-44431-4

% \bibitem[\protect\citeauthoryear{Treumann}{2006}]{Treumann2006} Treumann R.~A., 2006, A\&ARv, 13, 229. doi:10.1007/s00159-006-0001-y

\bibitem[\protect\citeauthoryear{Turner et al.}{2021}]{Turner2021} Turner J.~D., Zarka P., Grie{\ss}meier J.-M., Lazio J., Cecconi B., Emilio Enriquez J., Girard J.~N., et al., 2021, A\&A, 645, A59. doi:10.1051/0004-6361/201937201

\bibitem[\protect\citeauthoryear{Turnpenney et al.}{2018}]{Turnpenney2018} Turnpenney S., Nichols J.~D., Wynn G.~A., Burleigh M.~R., 2018, ApJ, 854, 72. doi:10.3847/1538-4357/aaa59c

\bibitem[\protect\citeauthoryear{Turnpenney et al.}{2020}]{Turnpenney2020} Turnpenney S., Nichols J.~D., Wynn G.~A., Jia X., 2020, MNRAS, 494, 5044. doi:10.1093/mnras/staa824

\bibitem[\protect\citeauthoryear{Varela et al.}{2016}]{Varela2016} Varela J., Reville V., Brun A.~S., Pantellini F., Zarka P., 2016, A\&A, 595, A69. doi:10.1051/0004-6361/201628607

\bibitem[\protect\citeauthoryear{Varela et al.}{2018}]{Varela2018} Varela J., R{\'e}ville V., Brun A.~S., Zarka P., Pantellini F., 2018, A\&A, 616, A182. doi:10.1051/0004-6361/201732091

\bibitem[\protect\citeauthoryear{Varela et al.}{2022}]{2022...659A..10V} Varela J., Brun A.~S., Strugarek A., R{\'e}ville V., Zarka P., Pantellini F., 2022, A\&A, 659, A10. doi:10.1051/0004-6361/202141181

\bibitem[\protect\citeauthoryear{Varela et al.}{2022}]{2022SpWea..2003164V} Varela J., Brun A.~S., Zarka P., Strugarek A., Pantellini F., R{\'e}ville V., 2022, SpWea, 20, e2022SW003164. doi:10.1029/2022SW003164


\bibitem[\protect\citeauthoryear{van der Holst et al.}{2014}]{vanderholst2014} van der Holst B., Sokolov I.~V., Meng X., Jin M., Manchester W.~B., T{\'o}th G., Gombosi T.~I., 2014, ApJ, 782, 81. doi:10.1088/0004-637X/782/2/81

\bibitem[\protect\citeauthoryear{van Haarlem et al.}{2013}]{vanHaarlem2013} van Haarlem M.~P., Wise M.~W., Gunst A.~W., Heald G., McKean J.~P., Hessels J.~W.~T., de Bruyn A.~G., et al., 2013, A\&A, 556, A2. doi:10.1051/0004-6361/201220873

\bibitem[\protect\citeauthoryear{Valenti \& Fischer}{2005}]{Valenti2005} Valenti J.~A., Fischer D.~A., 2005, yCat, 215. doi:10.26093/cds/vizier.21590141


\bibitem[\protect\citeauthoryear{Vedantham et al.}{2020}]{Vedantham2020} Vedantham H.~K., Callingham J.~R., Shimwell T.~W., Tasse C., Pope B.~J.~S., Bedell M., Snellen I., et al., 2020, NatAs, 4, 577. doi:10.1038/s41550-020-1011-9

\bibitem[\protect\citeauthoryear{Vidotto et al.}{2012}]{Vidotto2012} Vidotto A.~A., Fares R., Jardine M., Donati J.-F., Opher M., Moutou C., Catala C., et al., 2012, MNRAS, 423, 3285. doi:10.1111/j.1365-2966.2012.21122.x

\bibitem[\protect\citeauthoryear{Vidotto et al.}{2015}]{Vidotto2015} Vidotto A.~A., Fares R., Jardine M., Moutou C., Donati J.-F., 2015, MNRAS, 449, 4117. doi:10.1093/mnras/stv618

\bibitem[\protect\citeauthoryear{Vidotto et al.}{2023}]{Vidotto2023} Vidotto A.~A., Bourrier V., Fares R., Bellotti S., Donati J.~F., Petit P., Hussain G.~A.~J., et al., 2023, A\&A, 678, A152. doi:10.1051/0004-6361/202347237


\bibitem[\protect\citeauthoryear{Weber et al.}{2017}]{Weber2017b} Weber C., Lammer H., Shaikhislamov I.-F., Chadney J.-M., Erkaev N., Khodachenko M.~L., Griessmeier J.-M., et al., 2017, pre8.conf, 317. doi:10.1553/PRE8s317

\bibitem[\protect\citeauthoryear{Winn et al.}{2006}]{Winn2006} Winn J.~N., Johnson J.~A., Marcy G.~W., Butler R.~P., Vogt S.~S., Henry G.~W., Roussanova A., et al., 2006, ApJL, 653, L69. doi:10.1086/510528

\bibitem[\protect\citeauthoryear{Wood et al.}{2021}]{Wood2021} Wood B.~E., M{\"u}ller H.-R., Redfield S., Konow F., Vannier H., Linsky J.~L., Youngblood A., et al., 2021, ApJ, 915, 37. doi:10.3847/1538-4357/abfda5

\bibitem[\protect\citeauthoryear{Yadav \& Thorngren}{2017}]{Yadav2017} Yadav R.~K., Thorngren D.~P., 2017, ApJL, 849, L12. doi:10.3847/2041-8213/aa93fd

\bibitem[\protect\citeauthoryear{Yantis, Sullivan, \& Erickson}{1977}]{Yantis1977} Yantis W.~F., Sullivan W.~T., Erickson W.~C., 1977, BAAS, 9, 453

\bibitem[\protect\citeauthoryear{Yu, Zijlstra, \& Jiang}{2021}]{Yu2021} Yu B., Zijlstra A., Jiang B., 2021, Univ, 7, 119. doi:10.3390/universe7050119

\bibitem[\protect\citeauthoryear{Zarka \& Genova}{1983}]{ZarkaGenova1983} Zarka P., Genova F., 1983, Natur, 306, 767. doi:10.1038/306767a0

\bibitem[\protect\citeauthoryear{Zarka}{1998}]{Zarka1998} Zarka P., 1998, JGR, 103, 20159. doi:10.1029/98JE01323

\bibitem[\protect\citeauthoryear{Zarka et al.}{2001}]{Zarka2001} Zarka P., Treumann R.~A., Ryabov B.~P., Ryabov V.~B., 2001, Ap\&SS, 277, 293. doi:10.1023/A:1012221527425

\bibitem[\protect\citeauthoryear{Zarka, Queinnec, \& Crary}{2001}]{2001P&SS...49.1137Z} Zarka P., Queinnec J., Crary F.~J., 2001, P\&SS, 49, 1137. doi:10.1016/S0032-0633(01)00021-6

\bibitem[\protect\citeauthoryear{Zarka, Cecconi, \& Kurth}{2004}]{Zarka2004} Zarka P., Cecconi B., Kurth W.~S., 2004, JGRA, 109, A09S15. doi:10.1029/2003JA010260

\bibitem[\protect\citeauthoryear{Zarka}{2007}]{Zarka2007} Zarka P., 2007, P\&SS, 55, 598. doi:10.1016/j.pss.2006.05.045

\bibitem[\protect\citeauthoryear{Zarka, Lazio, \& Hallinan}{2015}]{Zarka2015} Zarka P., Lazio J., Hallinan G., 2015, aska.conf, 120. doi:10.22323/1.215.0120

\bibitem[\protect\citeauthoryear{Zarka et al.}{2018}]{Zarka2018} Zarka P., Marques M.~S., Louis C., Ryabov V.~B., Lamy L., Echer E., Cecconi B., 2018, A\&A, 618, A84. doi:10.1051/0004-6361/201833586

\bibitem[\protect\citeauthoryear{Zarka et al.}{2020}]{Zarka2020}
Zarka P., Girard J.~N., Tagger M., et al., 2020, in Proceedings of the International Conference on NenuFAR.~Title:The low-frequency radio telescope NenuFAR

\bibitem[\protect\citeauthoryear{Zarka et al.}{2021}]{Zarka2021}
Zarka P., Magalh\~aes F. P., Marques M. S., Louis C. K.,Echer E.,Lamy L., Cecconi B., Prangé R., 2021, JGR Space Physics, 126, e2021JA029780, doi: 10.1029/2021JA029780

\bibitem[\protect\citeauthoryear{Zarka et al.}{2025a}]{Zarka2025} Zarka P., Louis C.~K., Zhang J., Tian H., Morin J., Gao Y., 2025, A\&A, 695, A95. doi:10.1051/0004-6361/202450950

\bibitem[\protect\citeauthoryear{Zarka et al.}{2025b}]
{Zarka2025HoE} Zarka P., HoE, 1-19. 
doi=10.1007/978-3-319-30648-3\_22-3


\bibitem[\protect\citeauthoryear{Zhang et al.}{2023}]{Zhang2023} Zhang J., Tian H., Zarka P., Louis C.~K., Lu H., Gao D., Sun X., et al., 2023, ApJ, 953, 65. doi:10.3847/1538-4357/acdb77

\bibitem[\protect\citeauthoryear{Zhang et al.}{2025}]{Zhang2025} Zhang X., Zarka P., Girard J.~N., Tasse C., Loh A., Mauduit E., Mertens F.~G., et al., 2025, arXiv, arXiv:2506.07912. doi:10.48550/arXiv.2506.07912


\bibitem[\protect\citeauthoryear{Zic et al.}{2020}]{Zic2020} Zic A., Murphy T., Lynch C., Heald G., Lenc E., Kaplan D.~L., Cairns I.~H., et al., 2020, ApJ, 905, 23. doi:10.3847/1538-4357/abca90
\end{thebibliography}
\end{document}